# Molecular Precursor Chemistry Shapes Early-Stage Reaction Pathways in Nitrogen-Doped Carbon Nanotube Synthesis

*Narayan N. Som[1], Lukasz Nowicki[1], Tomasz Gizewski[2], Karolina Z. Milowska[3,4], Anna Wroblewska[5], Caghan Tuncer[6], Maciej Trzaskowski[1], Olga Iwasinska-Kowalska[7], Tomasz Plocinski[8], Mariusz Zdrojek[5],Krzysztof K. Koziol[6], Malgorzata Jakubowska[1,9], Agnieszka Lekawa-Raus[1]**

[1] Centre for Advanced Materials and Technologies (CEZAMAT), Warsaw University of Technology, Warsaw, Poland;

[2] Lublin University of Technology, Faculty of Electrical Engineering and Computer Science, Lublin, Poland

[3]BCMaterials, Basque Center for Materials, Applications and Nanostructures, UPV/EHU Science Park, 48940 Leioa, Spain

[4]Ikerbasque, Basque Foundation for Science, Bilbao, Spain

[5] Faculty of Physics, Warsaw University of Technology, Warsaw, Poland

[6]Cranfield University, Cranfield, United Kingdom

[7]Faculty of Mechatronics, Warsaw University of Technology

8Faculty of Materials Science and Engineering

[9] Faculty of Mechanical and Industrial Engineering, Warsaw University of Technology, Warsaw, Poland

*Corresponding author

E-mail address: agnieszka.raus@pw.edu.pl (A. Lekawa-Raus)

**Abstract**

The structural characteristics of carbon nanotubes are usually discussed in terms of processes occurring during nanotube nucleation and subsequent growth from an already formed catalyst nanoparticle. Here, we show that their structural trajectory can begin much earlier, within the first nanoseconds of precursor decomposition and catalyst formation. We combine experiment and reactive molecular dynamics to compare nitrogen-free CNT synthesis with two closely related aromatic nitrogen precursors, pyridine and pyrazine, containing one and two nitrogen atoms per molecule, respectively. Experimentally, both nitrogen-containing precursors promote bamboo-like CNT formation and produce substantially more structurally disordered CNTs than the nitrogen-free reference. Notably, nitrogen incorporation does not scale simply with the amount of nitrogen supplied, demonstrating a strong contribution from precursor chemistry. Reactive molecular dynamics reveals that nitrogen-containing precursors significantly modify local Fe–carbon and Fe–$SiO_2$ interactions during catalyst evolution, while pyridine and pyrazine exhibit distinct decomposition pathways, with pyrazine showing significantly greater decomposition within the simulated timescale. These results demonstrate that precursor chemistry influences the local chemical environment and early catalyst evolution, highlighting its potential role in the initial stages of CNT growth.

## 1. Introduction

Nitrogen-doped carbon nanotubes (N-CNTs) are a special class of CNTs widely studied as chemically active, electrically conductive one-dimensional scaffolds for a range of applications, including electrocatalysis, energy-storage devices (Li–S, Zn–air, and Na-ion batteries, as well as supercapacitors), electrochemical and optical sensors, biofuel cells and bioelectrodes, electromagnetic-wave absorption and EMI shielding, and metal-free catalysis [1–8]. Nitrogen incorporation modifies the local electronic structure and charge distribution, creates active sites, and enhances both reactant adsorption and CNT wettability, thereby promoting electrochemical reactions [1,9,10].

Nitrogen incorporation into CNTs can be achieved through two general approaches: in situ doping during CNT growth and post-synthesis functionalization or doping[11–15]. Of these, in situ synthesis offers several important advantages. Nitrogen atoms can be incorporated throughout the nanotube walls, enabling relatively high nitrogen contents, which can exceed 5 at.% under optimized growth conditions [13,15]. This approach can provide a more homogeneous distribution of nitrogen throughout the material while avoiding additional post-synthesis treatment. Methods such as floating-catalyst chemical vapor deposition (FC-CVD) further enable scalable, one-step production of large quantities of N-CNTs [16,17].

Because nitrogen incorporation occurs during nanotube growth, it can also influence the resulting nanotube morphology, often promoting the formation of characteristic bamboo-like structures. In bamboo-shaped N-CNTs, internal graphitic partitions and highly curved regions introduce a high density of edge-plane-like and chemically active sites along the nanotube walls and can also facilitate the confinement of catalytic nanoparticles, features that are particularly advantageous in electrochemical and catalytic applications [9,14,18–22].

Although the synthesis of N-CNTs have been widely reported, the mechanism governing their formation remains poorly understood. In this work, we investigate N-CNT growth by FC-CVD using a controlled comparison of two well-established nitrogen-containing precursors, pyridine and pyrazine. Pyridine contains one nitrogen atom, whereas pyrazine contains two nitrogen atoms within an aromatic ring, resulting in differences in molecular stability, decomposition pathways, and interactions with transition-metal surfaces [15,23]. Experimental studies have demonstrated that both precursors can promote nitrogen-containing and bamboo-like CNT structures [15,24–27]. However, pyridine and pyrazine have predominantly been investigated separately and under different synthesis conditions, making it difficult to distinguish the effects of precursor chemistry from those of temperature, catalyst composition, precursor concentration, and other process parameters [18,28–30]. A direct side-by-side comparison of their behaviour during in situ CNT synthesis under otherwise identical CVD conditions is therefore lacking.

At the same time, recent advances in machine-learning-driven molecular dynamics have substantially expanded the ability to model CNT growth at the atomistic level. Near-microsecond simulations have now enabled direct observation of CNT nucleation, growth, defect formation and healing on metal catalysts, while subsequent studies have addressed chirality-dependent growth kinetics, catalyst–CNT edge dynamics and catalyst-cluster evolution [31,32]. However, these approaches predominantly focus on the later stages of CNT growth, typically starting from pre-formed metal catalyst nanoparticles and simplified carbon supply rather than explicitly resolving the molecular chemistry that precedes nanotube nucleation. For example, the DeepCNT-22 framework models SWCNT growth from atomic carbon supplied to a pre-formed Fe catalyst [31–34]. Consequently, the coupled processes of molecular precursor decomposition, catalyst formation, and interaction of precursor-derived species with the evolving catalyst remain much less explored.

In our recent work, we initiated this line of investigation by introducing a precursor-resolved reactive MD approach to study early-stage FC-CVD chemistry, explicitly modelling hydrocarbon and ferrocene decomposition, Fe-cluster nucleation, and substrate interactions prior to CNT nucleation [35]. Building on this framework, the present study extends the approach to nitrogen-assisted CNT synthesis and combines it with controlled FC-CVD experiments. Under fixed baseline conditions of 760 °C, an Ar flow rate of 1 L $min^{-1}$, and 5 wt% ferrocene, CNT synthesis from pure toluene was compared with synthesis in the presence of pyrazine or pyridine. CFD–thermal simulations were used to assess whether the addition of nitrogen-containing precursors alters the reactor-scale flow and temperature fields, while SEM, TEM, Raman spectroscopy, and EDX were employed to characterize the resulting CNT morphology, structural disorder, and nitrogen content. Reactive molecular dynamics simulations were then used to investigate the corresponding early-stage chemistry, explicitly including toluene, pyridine or pyrazine, ferrocene, argon, and the $SiO_2$ substrate to resolve precursor decomposition and interactions with the evolving Fe catalyst.

## 2. Methods

### 2.1 Experimental procedure

A custom-designed horizontal floating catalyst chemical vapor deposition (FC-CVD) reactor was employed for the synthesis of carbon nanotube (CNT) arrays. The reactor system consisted of a high-purity quartz tube positioned within a tubular furnace featuring three independently controlled heating zones. Each zone was regulated by a dedicated temperature controller, enabling the generation of programmable thermal gradients along the reactor length. To facilitate efficient vaporization of the precursor solution, both the syringe containing the feedstock and the Argon carrier gas line were preheated to 70 °C prior to injection into the reactor. A liquid precursor mixture containing the carbon source and ferrocene as the catalyst precursor was continuously supplied at a flow rate of 5.6 mL $h^{-1}$ using a precision syringe pump. The solution was introduced into the reactor through a stainless-steel capillary and transported by an Argon flow of 1 L $min^{-1}$ (99.999% purity).

### 2.2 FEM modeling

In this study, a coupled CFD–thermal FEM model of a tubular CVD reactor is used to describe fluid flow and heat transfer governing precursor transport, residence time, and temperature distribution. The strong coupling between convective and thermal fields is resolved within a unified multiphysics framework.

The reactor comprises an outer shielding tube and a coaxial process tube (Fig. S3), whose inner wall defines the deposition surface $\Gamma_{su\beta}$. The precursor gas flows through the annular domain $\Omega_{ann}$, where transport phenomena develop under axisymmetric conditions, leading to thermal and hydrodynamic boundary layers and radial–axial temperature gradients due to indirect heating. The problem is formulated as a 2D axisymmetric model in cylindrical coordinates (r,z) [36,37].

### 2.3 Material characterisation

The materials obtained were compared using a scanning electron microscope (SEM) SU8230 instrument (Tokyo, Japan) from Hitachi and Spectra 200 (S)TEM microscope from Thermo Scientific. Thermogravimetric analysis of the samples was carried out using a Q500 thermogravimetric analyzer from TA Instruments (USA). The decomposition was carried out surrounded by synthetic air with a flow rate of 100 ml/min. The temperature rise was 5°C/min and the final temperature was 800 °C. Raman Spectroscopy was performed on an inVia Raman Microscope from Renishaw (UK). Each sample was measured 100 times, and the average results are shown.

### 2.4 Computational Details of Atomistic Modeling

We performed classical reactive molecular dynamics (RMD) simulations using the ReaxFF framework implemented in LAMMPS [38,39] to investigate the thermal decomposition behavior of ferrocene and its interaction with hydrocarbon feedstocks under FCCVD-like conditions. In this study, we employed a hybrid ReaxFF parameter set to accurately capture the multicomponent chemistry involving Fe, C, H, O, Si, and N. The reactive force field (ReaxFF) parameters employed in this study were adopted from established literature sources to ensure a consistent and physically reliable description of all relevant interactions. Specifically, the Fe/C/O/H/Si parameter set was taken from the ReaxFF formulation reported by Zou et al.[40], as distributed within the standard LAMMPS ReaxFF framework. To incorporate nitrogen chemistry, the C/H/O/Si/N interaction parameters were adopted from ReaxFF for C H O Si N systems, which provides a validated description of nitrogen interactions compatible with hydrocarbon and oxide environments [41]. In addition, Fe–N cross-interaction parameters were obtained from the thesis *"Classical and ReaxFF Molecular Dynamics Simulations of Fuel Additives at the Solid–Fluid Interface"*, enabling an explicit treatment of metal–nitrogen interactions [40–42]. The ferrocene decomposition, iron clustering, and hydrocarbon breakdown over substrate were carried out at 760 °C. We have carried out the simulation with 760 °C providing favorable conditions from experimental observation for N-doped CNT.

## 3. Results and discussion

The morphology and structural characteristics of carbon nanotubes (CNTs) are strongly influenced by the nature of the precursor composition during growth [35,43]. In this study, particular emphasis was placed on investigating the effect of nitrogen-containing compounds on CNT structure and identifying the early-stage chemical pathways involved in N-CNT formation[18,44,45]. To this end, pyrazine and pyridine were introduced as nitrogen sources into a toluene-based feedstock along with ferrocene source based catalysis, while maintaining consistent synthesis conditions. All experiments were conducted at 760 °C using toluene as the carbon precursor and ferrocene (5 wt%) as the catalyst source. The nitrogen-containing additives were added to the base feedstock at 5 wt% pyrazine, 10 wt% pyrazine, or 10 wt% pyridine (samples denoted as T760C_Pz5%_Fer5%, T760C_Pz10%_Fer5%, and T760C_Pd10%_Fer5%, respectively). The reference samples containing only toluene and ferrocene are denoted as T760C_Fer5%.

The TEM images (**Figure 1**) reveal that both pyrazine and pyridine promote the formation of a bamboo-type morphology, which has not been observed in the reference samples. A detailed microscopic characterization of the reference CNTs has been reported in our previous work [35]. As shown in Figures 1a, c, and e for the 10 wt% pyrazine sample (T760C_Pz10%_Fer5%), the nanotubes exhibit pronounced internal compartments, with periodic dark contrasts along the tube axis associated with encapsulated catalyst particles and graphitic closures (Figure 1a).

This compartmentalization becomes more evident in Figure 1c, where the internal structure displays clearly separated graphitic sections, indicative of repeated closure and re-initiation of growth. Furthermore, Figure 1e shows a relatively large, elongated catalyst particle confined within the tube core, while the clearly resolved concentric graphitic walls confirm the multi-walled nature of the nanotube.

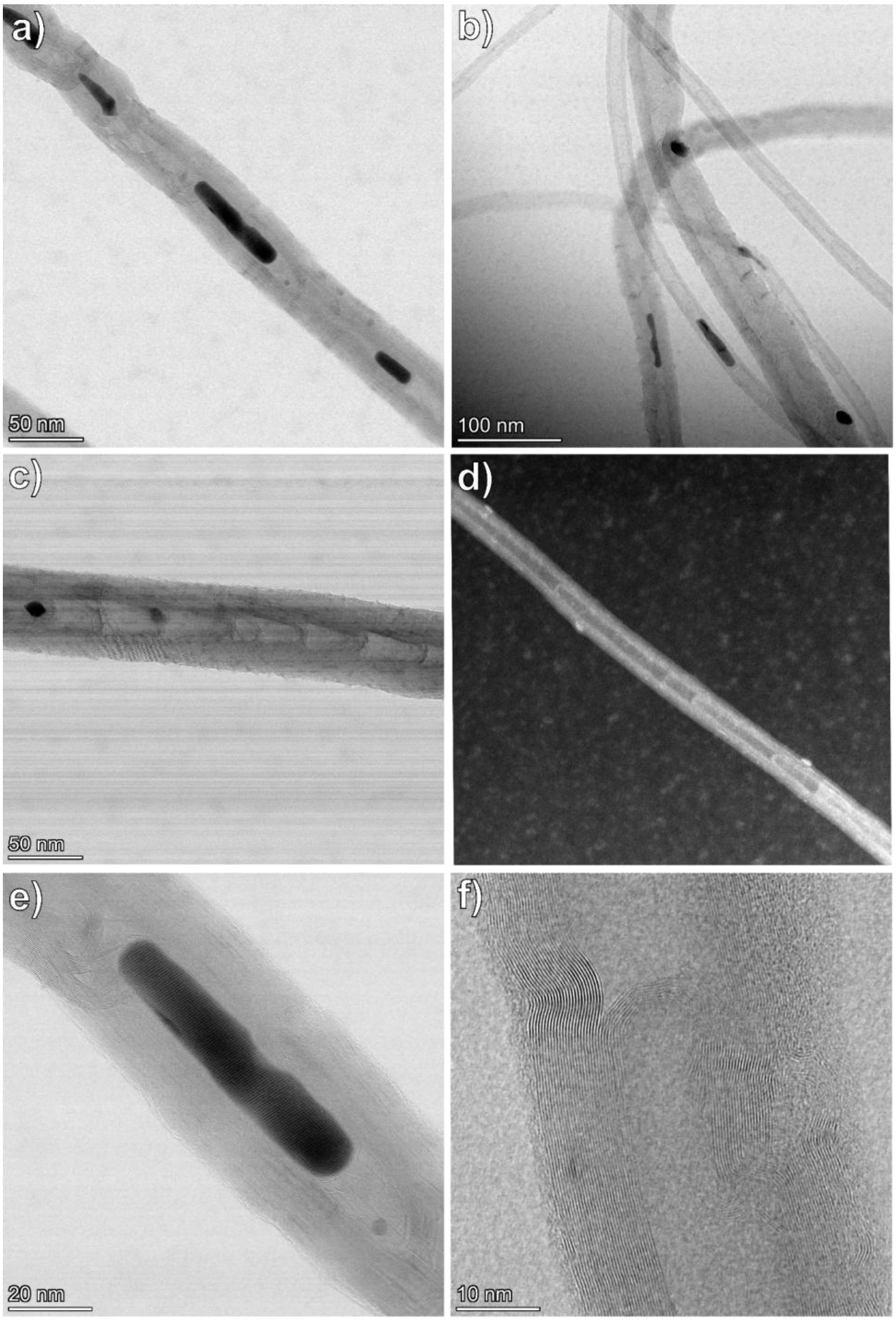

**Figure 1:** TEM images of bamboo-shaped carbon nanotubes obtained with a), c), e) 10 wt.% pyrazine and b), d), f) 10 wt.% pyridine

The pyridine-derived carbon nanotubes (T760C_Pd10%_Fer5%, **Figure 1b, d, f**) also demonstrate bamboo-like features, but with comparatively less regularity in compartment spacing and wall definition. **Figure 1b** shows multiple nanotubes with intermittent segmentation, though the compartments appear less uniformly distributed than in the pyrazine case (T760C_Pz10%_Fer5%). In **Figure 1d**, the nanotube walls appear more continuous with fewer clear closures, indicating a lower degree of structural compartmentalization. High-resolution imaging in **Figure 1f** reveals graphitic layers with some degree of ordering, but with noticeable distortions and defects, likely associated with nitrogen incorporation.

Overall, while both precursors facilitate the formation of bamboo-like CNTs, pyrazine appears to promote more regular and well-defined compartmental structures, whereas pyridine results in relatively less ordered segmentation, reflecting differences in nitrogen incorporation and its influence on growth dynamics.

To substantiate the presence and extent of structural modification, complementary EDX and Raman spectroscopic characterisations were performed. Energy-dispersive X-ray spectroscopy (EDX) was employed to evaluate the elemental composition and nitrogen content of the synthesized materials (**Figure 2**). The EDX results revealed that for the 5% pyrazine sample, the nitrogen content was low, reaching a mass concentration of only 0.38 wt%. In contrast, increasing the pyrazine concentration to 10% resulted in a substantial rise in nitrogen content, reaching 3.68 wt%. For the sample synthesized with 10% pyridine, an intermediate mass concentration of 1.80 wt% nitrogen was recorded. Taking into account that 5 wt% pyrazine supplies a comparable amount of nitrogen to 10 wt% pyridine, these findings clearly indicate that final nitrogen content depends strongly on both precursor concentration and precursor chemistry.

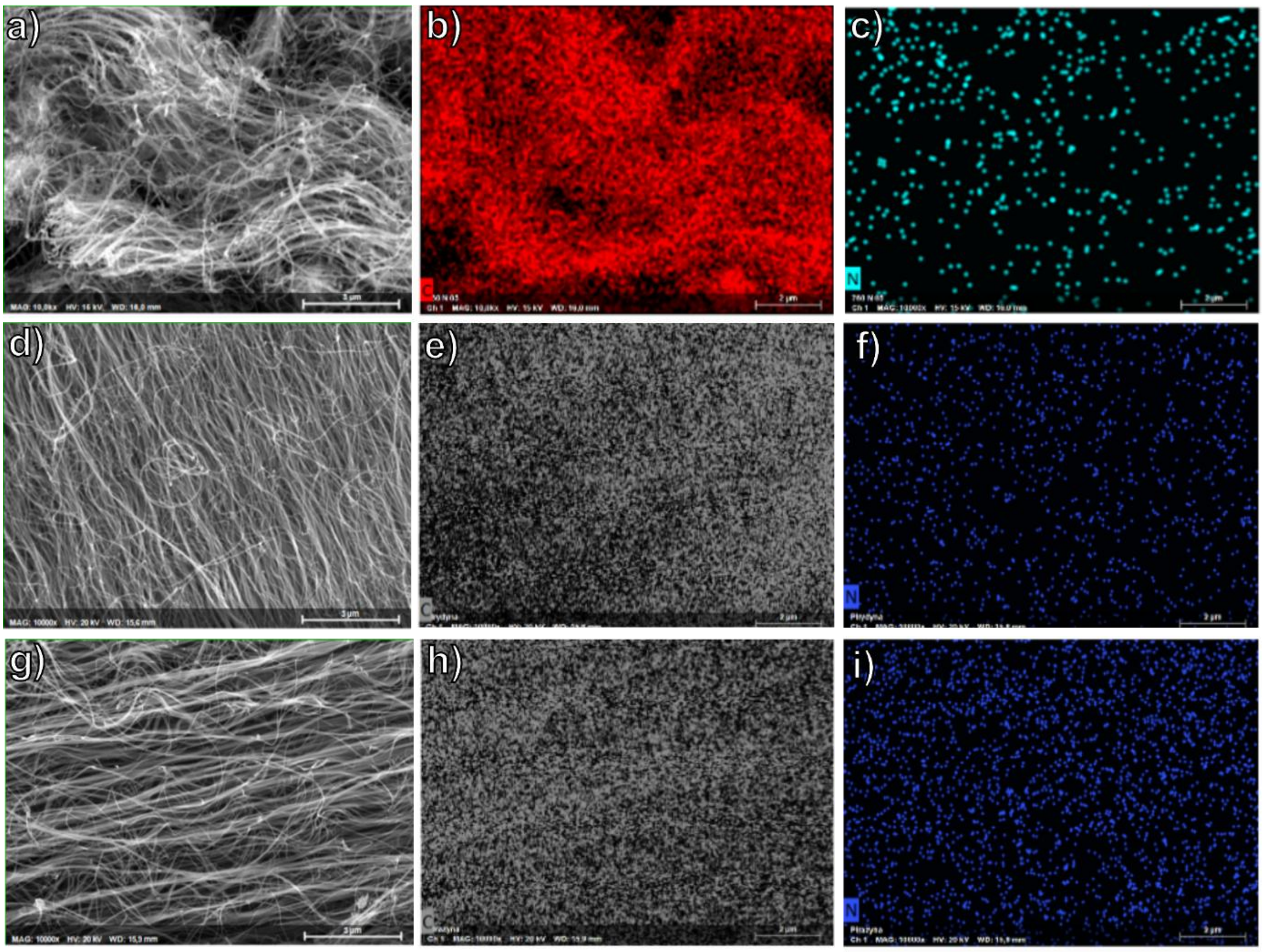

**Fig. 2** EDX mapping analysis of nitrogen-doped carbon nanotubes synthesized with different precursors and concentrations: (a–c) 5% pyrazine-derived sample, (d–f) 10% pyrazine-derived sample, and (g–i) 10% pyridine-derived sample. Panels (a), (d), and (g) show the overall morphology/SEM-related view or mixed elemental maps, while panels (b, e, h) and (c, f, i) present individual elemental mappings corresponding to carbon (C) and nitrogen (N) distributions, respectively.

Nitrogen incorporation is expected to induce discernible Raman spectral changes, including shifts in the G, D, and 2D bands, as well as an increased defect density, reflected in variations of the $I_D/I_G$ ratio ratio. For both pristine and nitrogen-doped nanotubes, five independent samples were analyzed, yielding over 100 Raman maps per material and more than 600 individual measurement points in total. The distributions of the G, D, and 2D peak positions, together with the $I_D/I_G$ ratios are presented as histograms in **Figure 3** for pristine CNTs, 10% pyrazine, and 10% pyridine, while the histograms for 5% pyrazine are provided in the Supplementary Information (SI). The Shapiro–Wilk test indicated that the distributions of the G, D, and 2D band positions, as well as the $I_D/I_G$ ratio, deviated from normality for all materials ($p < 0.05$). Consequently, the non-parametric Kruskal–Wallis test was applied, revealing statistically significant differences among at least one of the synthesized materials for each of the four evaluated parameters ($p < 0.05$).

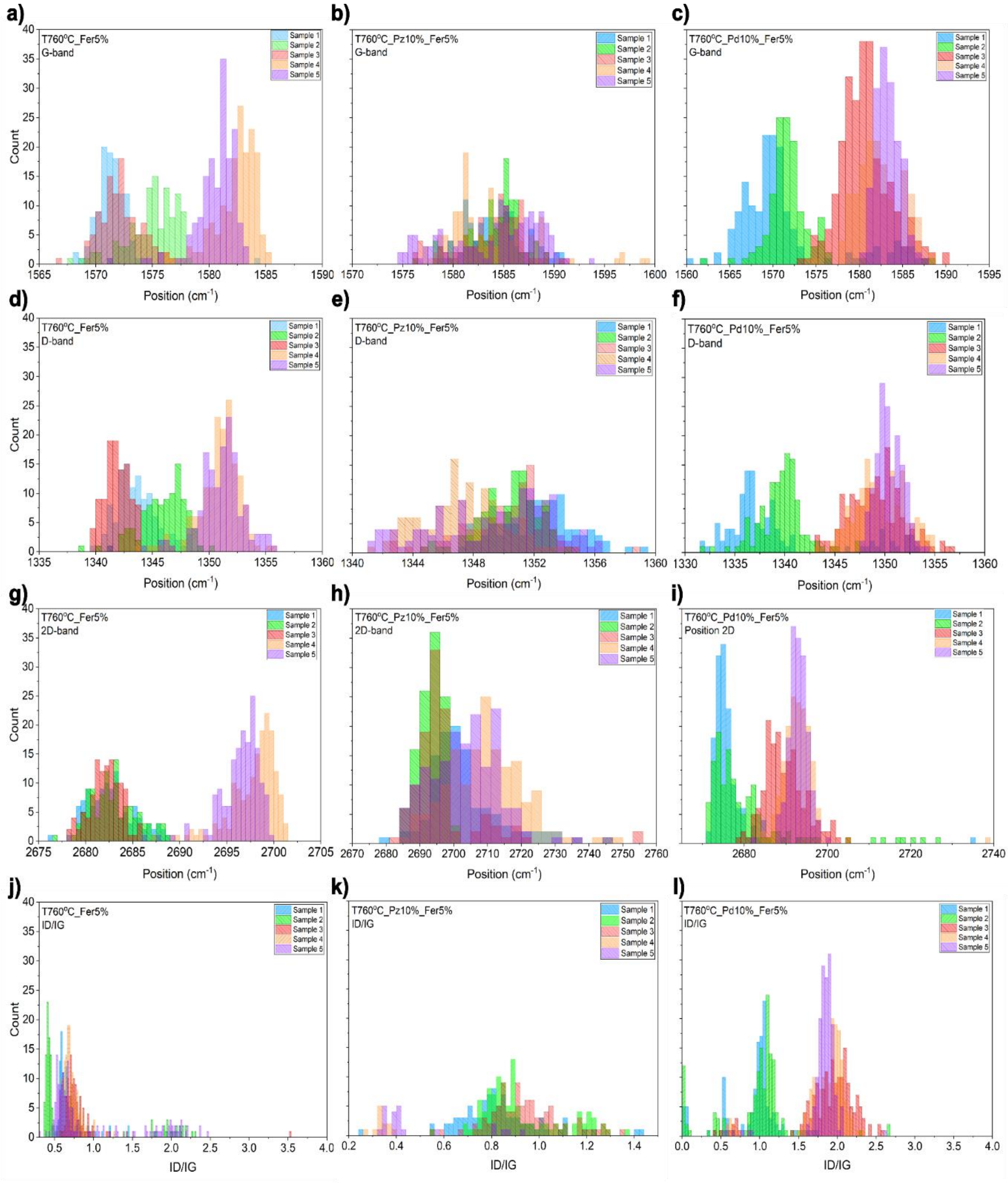

**Fig 3:** Raman shifts in the G, D, and 2D bands, together with changes in the ratio: comparison between pristine CNTs (a, d, g, j), pyrazine-derived CNTs (b, e, h, k), and pyridine-derived CNTs (c, f, i, l).

To further resolve pairwise differences, a Bonferroni post hoc correction was implemented across all groups. For the 5 wt% pyrazine sample, statistically significant differences were observed for the $I_D/I_G$ ratio and the 2D band, while the G and D band positions did not differ significantly from those of the pristine CNTs. The 2D band exhibited only a small upshift, accompanied by a modest increase in $I_D/I_G$, consistent with the low nitrogen content detected by EDX. For the 10 wt% pyrazine sample, statistically significant changes were observed for all three Raman bands and the $I_D/I_G$ ratio. The G, D, and 2D bands all exhibited upshifts. Such simultaneous upshifts have previously been associated with p-type doping of CNTs and may therefore indicate an acceptor-like electronic contribution [46]. Acceptor-like behaviour has also been reported for pyridinic N configurations [47]. At the same time, $I_D/I_G$ increased from approximately 0.1–0.5 in the pristine material to around 1.0, indicating an increased degree of structural disorder, consistent with the presence of defect-associated nitrogen configurations. In contrast, the pyridine-derived CNTs exhibited statistically significant changes in the D and 2D band positions and in the $I_D/I_G$ ratio, while the G-band position did not change significantly. Both the D and 2D bands exhibited downshifts, which may be consistent with an n-type electronic contribution previously associated with substitutional N doping [47]. Simultaneously, the pyridine-derived CNTs showed considerably higher $I_D/I_G$ values, reaching approximately 2.0–2. 5, indicating a substantially higher degree of structural disorder [45].

The partial substitution of carbon atoms by nitrogen, associated with the formation of new C–N bonds and the concomitant increase in defect density, suggests that the modified nanotube structures may exhibit reduced thermal stability and undergo decomposition at lower temperatures compared to pristine counterparts. To verify this hypothesis and provide a more comprehensive comparison between undoped and nitrogen-doped nanotubes, thermogravimetric analysis (TGA) was performed. The corresponding results are presented in **Figure 4a.**

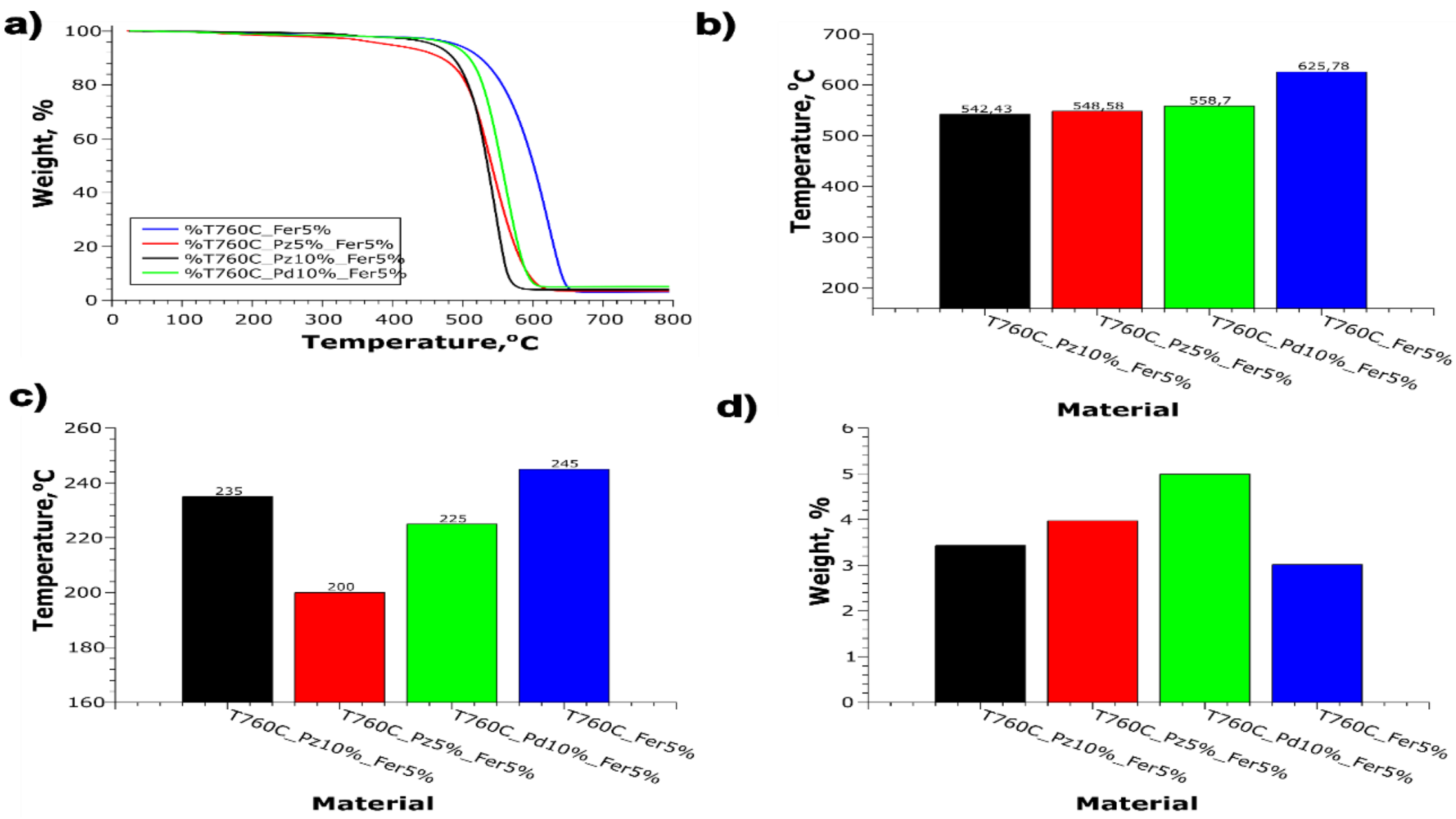

**Fig. 4:** **a)** Thermogravimetric Analysis for pure and doped CNTs **b)** DTG peak position corresponding to sp²-carbon decomposition **c)** Width of DTG peak (Tmax-Tmin), **d)** Catalyst-derived residue for each sample

For all the materials obtained, a slight decrease in mass was observed in the initial phase, indicating the evaporation of various species adsorbed from the environment, such as water [48]. Further weight loss is caused by the decomposition of carbon forms with a weak graphite structure, such as amorphous carbon, which occurs for individual samples in the following ranges: for samples T760C_Pz10%_Fer5%, T760C_Pz5%_Fer5%, and T760C_Pd10%_Fer5% – 100°C – 400°C, and for the reference sample T760C_Fer5% – 100°C – 440°C [46]. The difference between these ranges may be due to differences in precursor composition and the resulting carbon structure. At elevated temperatures, the dominant mass loss is observed, corresponding to the decomposition of more ordered sp2-hybridized carbon structures, including both pristine and nitrogen-doped carbon nanotubes. For materials synthesized with nitrogen-containing precursors (pyrazine and pyridine), this decomposition occurs within the temperature range of approximately 440–640 °C. In contrast, for the reference sample (T760C_Fer5%), the corresponding decomposition extends to a higher temperature, reaching up to 685 °C. The lower oxidation temperatures of the nitrogen-containing samples are consistent with their higher degree of structural disorder observed by Raman spectroscopy. Consequently, the nitrogen-doped nanotubes exhibit lower thermal stability and an increased susceptibility to oxidative degradation compared with their pristine counterparts. To provide a more detailed evaluation of these effects, derivative thermogravimetric (DTG) curves were constructed, enabling precise determination of the position (**Figure 4b)** and width of the principal decomposition peaks (**Figure 4c**). The calculated peak widths of the obtained materials demonstrate that changes in the relative proportion of components influence the thermal profile. Furthermore, the pyridine-doped nanotubes exhibit a higher decomposition peak temperature and a narrower peak width compared with the pyrazine-doped counterpart, which may reflect differences in the nature of nitrogen incorporation. In particular, the higher thermal stability would be consistent with a greater contribution from substitutional/graphitic N configurations, whereas pyridinic N is generally associated with less thermally stable defect sites. The lowest residual catalyst content observed for pristine CNTs was 3.01% (**Figure 4d**). The introduction of 5 wt% and 10 wt% pyrazine resulted in a marginal increase in impurity levels, not exceeding 1% relative to the reference material. In contrast, the use of pyridine led to a more pronounced increase in catalyst-derived residues (4.99%) compared with the pyrazine-doped samples, indicating a higher degree of residual catalytic species within the final product. These results clearly demonstrate that the nature of the nitrogen source and its concentration are critical parameters governing the formation of nitrogen-doped carbon nanotubes (N-CNTs), directly influencing their thermal behaviour and structural characteristics.

### 3.1. Flow in the cylindrical reaction domain

The coupled CFD–thermal analysis (see calculation details in Supporting Information) shows that nitrogen addition does not significantly alter the global flow and temperature fields compared to the nitrogen-free case. Both 2D and 3D simulations yield similar velocity and thermal distributions, confirming the robustness of the numerical model.

A stable laminar flow regime persists across the main synthesis zone, extending from the middle of the first heating section to the end of the third. Within this region, velocity profiles remain smooth, with no indication of instability or transition to turbulence, ensuring overall hydrodynamic stability despite changes in gas composition (**Figure 5**).

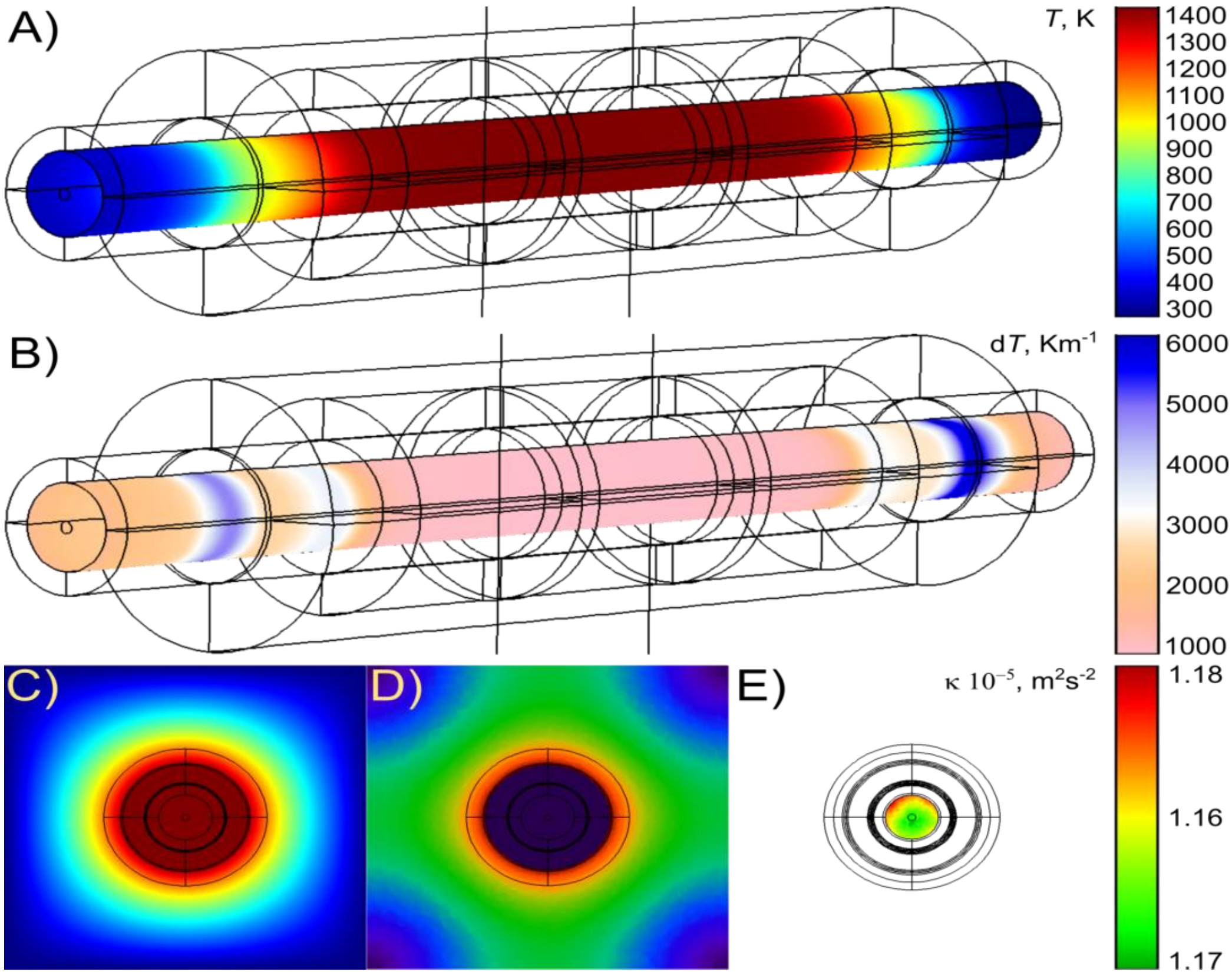

**Fig. 5**: Steady-state thermal and flow characteristics of the CVD reactor. (A) Three-dimensional temperature distribution within the reactor after reaching steady-state conditions. The central synthesis zone is characterized by a nearly uniform high temperature, whereas pronounced temperature variations occur mainly within the inlet and outlet transition regions. (B) Spatial distribution of the temperature-gradient magnitude under steady-state conditions. The highest temperature gradients are localized at the boundaries between the heated reaction zone and the cooler inlet and outlet sections, while the central synthesis region exhibits comparatively low thermal gradients. (C) Transverse temperature distribution in a representative cross-section of the reactor, showing an approximately axisymmetric thermal field. (D) Corresponding transverse distribution of the temperature-gradient magnitude, indicating that the strongest radial gradients occur near the reactor wall. (E) Cross-sectional distribution of the turbulence-

related coefficient, showing only minor spatial variations and confirming the hydrodynamic stability of the flow within the analyzed region.

Gravity induces only a slight asymmetry in pressure, mainly near the lower wall of the tube, while buoyancy-related effects cause weak variations in turbulence quantities. Overall, the flow and thermal fields remain stable and well-ordered, supporting reproducible nanostructure synthesis.

### 3.2 Atomistic modeling

Reactive molecular dynamics simulations were performed on α-SiO2 (101) slabs under experimental growth conditions (760 °C) to investigate the influence of nitrogen-containing precursors on Fe catalyst evolution and early-stage reaction chemistry leading to N-doped CNT formation. Two precursor-loading scenarios containing 13 and 30 ferrocene molecules (Fc13 and Fc30) were examined to evaluate the effect of Fe availability on catalyst nucleation and cluster evolution. The Fc13 systems were primarily used to analyze early-stage precursor decomposition, radical formation, and initial Fe segregation dynamics, while the Fc30 systems were employed to investigate extended catalyst evolution in the presence of nitrogen-containing species. Since both systems exhibited similar qualitative trends, the detailed atomistic decomposition analysis of the Fc13 systems is provided in the Supplementary Information, whereas the main manuscript focuses primarily on the Fc30 systems due to their closer relevance to experimentally observable CNT synthesis conditions.

The Fc13TAr simulations revealed that ferrocene rapidly decomposes near the quartz surface, releasing Fe atoms together with cyclopentadienyl-derived carbon fragments that subsequently evolve into short linear carbon chains. Simultaneously, toluene undergoes hydrogen abstraction and bond cleavage reactions, producing highly reactive hydrocarbon radicals such as methyl, benzyl, and small carbon species. In nitrogen-containing systems, pyrazine (Fc13Pz10TAr) and pyridine (Fc13TPd10Ar) decomposition generated low nitrogen-rich fragments (such as $C_2H_2$, $C3H3^+$or $C4H3^+$ rather than HCN radicals) and transient Fe–N interactions that modified local carbon chemistry and catalyst restructuring behavior. By performing 2 ns molecular dynamics simulations on a quartz slab at 760 °C, we obtained insights into the early-stage surface-assisted CNT growth process, including precursor decomposition, catalyst evolution, and Fe cluster formation. Within the simulated timescale, pyrazine (Fc13Pz10TAr) showed a tendency toward enhanced Fe and carbon aggregation compared with the nitrogen-free system, whereas pyridine (Fc13Pd10TAr) exhibited comparatively weaker surface interactions and less pronounced cluster evolution (**Figure S4**). These observations should be regarded as short-timescale trends, as longer simulations may be required to determine whether these differences persist during subsequent stages of CNT growth. These differences highlight that not only the nitrogen concentration but also the molecular structure of the nitrogen precursor plays a decisive role in determining Fe cluster behavior and, consequently, the CNT synthesis mechanism.

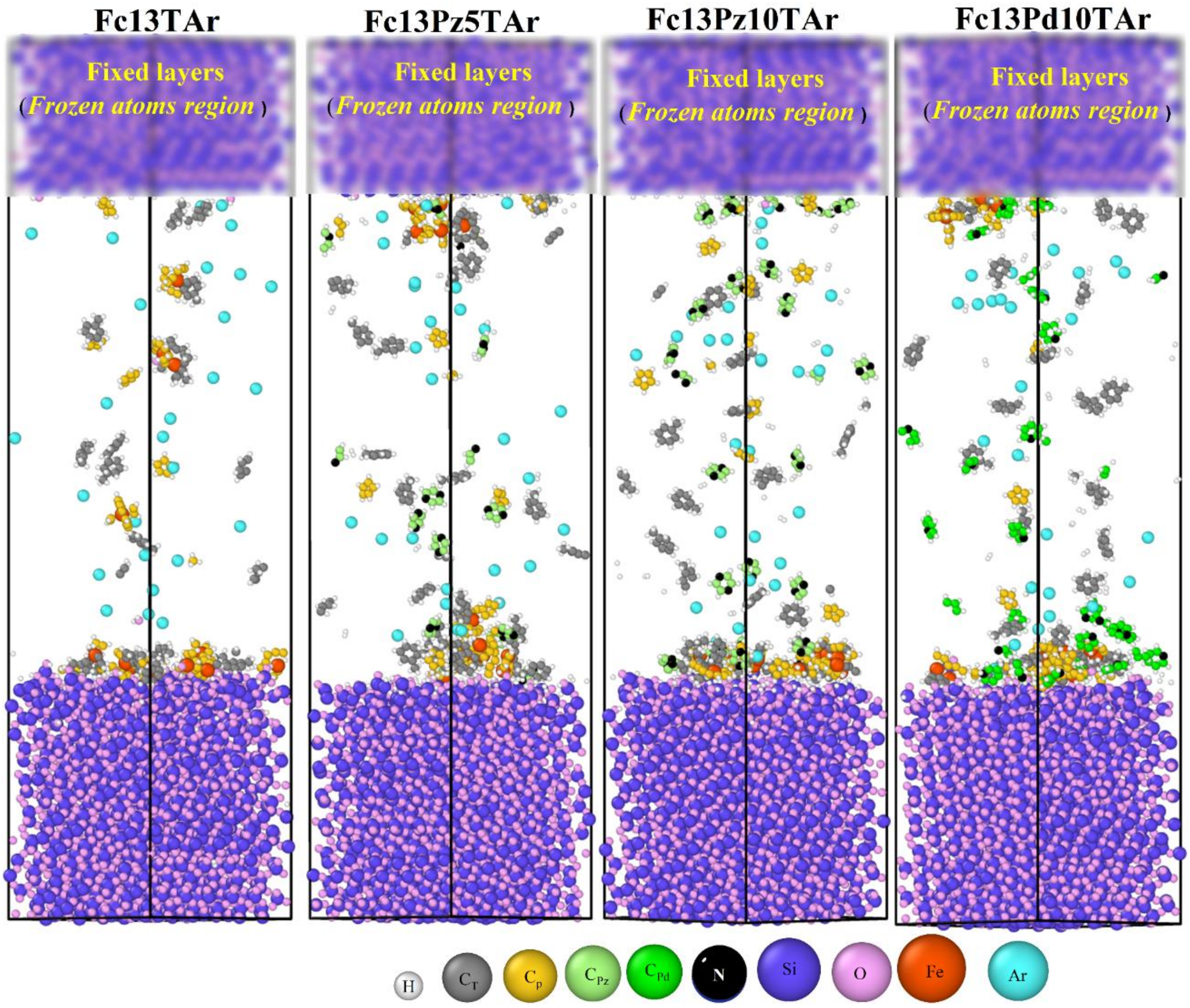

**Fig. 6:** Snapshot from the molecular dynamics simulation showing the radicals formed at the specified simulation time. Atoms are color-coded as follows: hydrogen (H, white), Toluene's Carbon (CT, grey), cyclopentadienyl carbon (C*p*, golden yellow), Pyrazine's carbon (CPz, pistachio green), pyridine's carbon (CPd, electric green), Silicon (Si, purple), oxygen (O, pink), iron (Fe, vibrant orange), Nitrogen (black) and argon (Ar, cyan)

Under nitrogen-free conditions, Fe atoms formed relatively stable clusters near the quartz surface, while carbon species largely remained as short-lived fragments with limited graphitic organization. In contrast, nitrogen-containing precursors significantly altered catalyst evolution and cluster stability. Pyrazine may promote faster Fe aggregation, enhanced formation of intermediate-sized Fe clusters, and increased carbon aggregation before 2ns (**Figure 6** and **Figure S5**). Pyridine exhibited comparatively weaker cluster evolution at 2ns initial and less pronounced segregation behavior, suggesting strong precursor-dependent catalyst dynamics. These observations suggest at the initial stage, nitrogen precursors may modify Fe cluster stability and restructuring behavior during early-stage CNT nucleation. Detailed decomposition pathways, radical evolution, and time-dependent cluster distributions are provided in the Supplementary Information.

### 3.2.1 Ferrocene Loading and Nitrogen-Rich Conditions Govern Early-Stage Fe-Cluster Evolution

For the system containing 13 ferrocene molecules (Fc13), the simulations qualitatively capture the influence of precursor chemistry on ferrocene decomposition and the subsequent evolution of Fe clusters, which is relevant to the early stages of CNT growth. However, to better understand the underlying mechanism, a detailed analysis of Fe-cluster aggregation, size, and compactness is required to determine how the precursor chemical environment influences catalyst evolution. In particular, we examine how the interactions among carbon species, nitrogen-containing precursors, and toluene influence early-stage precursor decomposition and the local chemical environment of the Fe catalyst under Fe-rich conditions, using 30 ferrocene molecules as the model system.

To address this, we performed an analysis including cluster analysis through the radius of gyration, radial distribution functions (RDF) for Fe–Fe and Fe–carbon interactions involving both toluene and nitrogen-based precursors, as well as charge analysis. These parameters play a crucial role in determining catalyst behavior and CNT growth at 760 °C. Based on the observations from the Fc13Pz5TAr system, the 5% pyrazine condition was not considered further for the Fe-rich Fc30 systems, as the Fc13Pz10TAr system exhibited more pronounced and compact Fe-cluster formation. Therefore, the 10% pyrazine concentration was selected for subsequent Fc30 simulations.

Therefore, to better understand the role of nitrogen-containing molecules, we further examined systems containing pyrazine and pyridine with the same number of molecules. This comparison provides a more consistent basis for assessing the influence of precursor chemistry and identifying trends in the early stages of CNT growth.

### 3.2.2 Fe-Cluster Evolution on the $SiO_2$ Surface under Fe-Rich Conditions

Thereafter, focus was extended to systems containing 30 ferrocene molecules in order to further understand Fe cluster formation on the slab surface and to examine the correlation between cluster formation and the presence of nitrogen-containing precursors. For this purpose, three systems were investigated at 760 °C, namely Fc30TAr, Fc30Pz10TAr, and Fc30Pd10TAr. This comparative approach allows us to evaluate how increased Fe precursor loading, together with different nitrogen sources, influences cluster aggregation behavior, and the stabilization of specific Fe cluster sizes relevant to early-stage CNT growth synthesis up to 4ns.

The density profile clearly shows the confined α-$SiO_2$ layers at the top (fixed and periodic image of Z-axis label as frozen atoms region and with blur effect in **Figure 7a**) and at bottom is mobile layer), while the reactive species evolve within the interfacial region **( See Figure 7b**).The Fe atoms exhibit stronger aggregation toward the slab surface, forming more compact and stabilized clusters at 4ns (box in **Figure 7a** and cluster analysis of **Figure 7d**) compared to the Fc13TAr system, where Fe species remain more dispersed (**Figure S5** and **Figure 6**).The hydrogen density is particularly high within the lower region of the quartz substrate (10–20 Å; **Figure 7b**), indicating the penetration of hydrogen molecules into the slab. Concurrently, slight degradation of the upper quartz surface is observed, accompanied by the formation of $H_2O$, visible as pink spheres within the simulation box (**Figure 7a**). These changes are likely associated with the interaction of hydrogen molecules with the passivated surface and surface oxygen species, consistent with observations reported in previous studies [49]. The charge density analysis (**Figure 7c**) reveals a clear redistribution of electronic charge across on the bottom of the slab, comparing top layer and bottom most layer (**Figure 7c**).

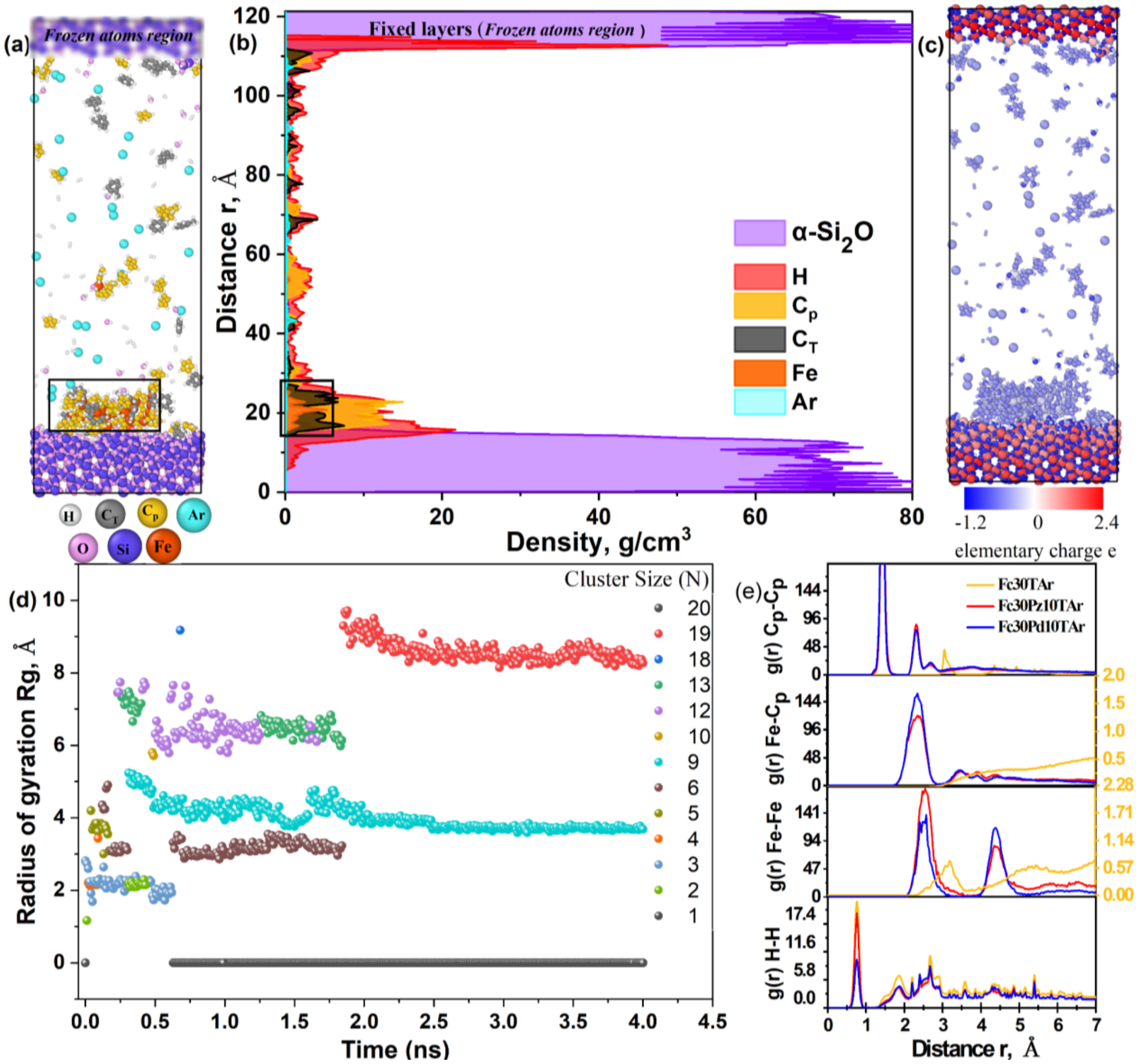

**Fig. 7:** Molecular dynamics post-processing results based on the first 5 ps extracted from the 4 ns simulation of the Fc30TAr system at 760 °C (1033 K): (a) atomic structures, (b) charge

distribution profiles, (c) charge profiles, and (d) cluster analysis. (e) Radial distribution functions, *g(r)*, for H–H, Fe–Fe, Fe–$C_p$, and $C_p$–$C_p$ pairs for the Fc30TAr, Fc30Pz10TAr, and Fc30Pd10TAr systems.

The slab surface initially exhibits a predominantly positive character, as indicated by the red-shaded regions in the bottom most layers of **Figure 7c**. Following interaction with the adsorbed and segregated species, a clear redistribution of charge is observed near the surface, reflected by the reduced intensity of the red regions and the emergence of negatively charged regions around the surface-bound species. This behavior suggests appreciable electronic interaction and charge redistribution between the slab and the adsorbed species. The Fe clusters remain nearly neutral, with a slight tendency toward negative charge relative to carbon atoms. In contrast, some carbon atoms, particularly those located in the middle regions of the section around 120 Å to 60 Å , exhibit an almost neutral charge state. This localized charge neutrality can act as a precursor to defect formation, which may play a significant role in influencing the subsequent growth and structural evolution of carbon nanotubes.

In the Fc30TAr system, cluster analysis (**Figure 7d**) reveals the formation and evolution of Fe clusters over time. At the initial stage (up to 0.5 ns), small clusters consisting of 3, 5, 9, and 13 atoms are observed. As the simulation progresses to 4 ns, the 9-atom cluster appears to be the most stable configuration, while the 13-atom cluster shows enhanced stability particularly between 1.5 and 2.0 ns. Subsequently, these clusters undergo reconstruction to form a more compact Fe cluster of 19 atoms. The radius of gyration analysis further confirms enhanced cluster compactness over a period of 4ns (listed in Table S2). Larger Fe clusters maintain relatively constant radii after formation, reflecting structural stability and tighter Fe–Fe interactions. This observation is consistent with the Fe–Fe radial distribution function (See **Figure 7e**, which displays sharper and more intense first neighbor peaks, indicating improved clustering and having slight atomic dispersion rather than compact. The Fe–$C_p$ and $C_p$–$C_p$ radial distribution functions largely gaussian in nature, shortening the Fe–$C_p$ distance, suggesting marginal strengthening of Fe–carbon interactions with major restructuring of carbon species of cyclopentadienyl (majorly yellow sphere observed near Fe cluster as show with drawn box in **Figure 7a)**. The hydrogen related radial distribution functions reveal a pronounced difference between systems. The H–H g(r) indicates higher hydrogen molecule formation whereas hydrogen production is significantly higher compared to the presence of pyrazine and pyridine precursors. Quantitative analysis shows that trapped hydrogen species increase steadily to a saturation value of 9 within 4 ns. This clearly demonstrates hydrogen infiltration into the substrate (can confirm its Density plot, **Figure 7b**), likely by limiting hydrogen radical stripping during toluene decomposition and reducing their mobility toward the surface. It modulates the local thermodynamic environment, enhances Fe cluster compactness and stability, suppresses excessive hydrogen penetration into the quartz substrate, and prevents surface degradation with 4ns simulation.

### 3.2.3 Influence of Nitrogen Precursors on Early-Stage Fe-Cluster Evolution

In Fc30TAr, the decomposition of ferrocene resulted in the aggregation of Fe atoms and subsequent cluster formation. Along with catalyst formation, the decomposition of ferrocene and toluene also produced carbon species that assembled into carbon chains with a maximum length of approximately 13 atoms, indicating the availability of reactive carbon fragments for the nucleation of carbon nanotubes.

To further investigate the influence of nitrogen-containing precursors on catalyst evolution and carbon growth mechanisms, we introduced pyrazine, say Fc30Pz10TAr. Compared to the Fc30TAr system, the presence of pyrazine significantly enhances the aggregation of Fe atoms and promotes more pronounced cluster formation. **Figure 8a** shows the optimized configuration obtained after 4 ns of simulation. Two regions of interest were defined in **Figure 8a** to examine the spatial evolution of Fe clusters. The analysis reveals a clear spatial differentiation in cluster size: larger Fe clusters preferentially form above the fixed α-$SiO_2$ slab, whereas smaller clusters remain dispersed and surface-bound. This coexistence of differently sized Fe clusters within the catalyst environment suggests heterogeneous nucleation sites and may contribute to the broad distribution of CNT diameters observed experimentally.

A similar structural trend is observed in the pyridine-functionalized system (**Fc30Pd10TAr**), which closely resembles the behavior of **Fc30TAr**. In both systems, Fe-cluster aggregation preferentially occurs near the α-$SiO_2$ slab surface, as highlighted by the inset box in **Figure 9a**. This consistent behavior suggests that variations in precursor composition and concentration significantly modulate the thermodynamics of cluster formation, thereby altering the spatial positioning and structural distribution of the Fe catalysts.

The spatial density distribution of carbon species provides further insight into the decomposition pathways of the precursors. The density profile (**Figure 8b**, Fc30Pz10TAr) exhibits a pronounced carbon peak in the region between 20 and 40 Å above the slab surface, which primarily corresponds to carbon atoms originating from the decomposition of toluene. Additionally, the larger Fe clusters are predominantly surrounded by carbon species derived from toluene near the surface, whereas cyclopentadienyl fragments originating from ferrocene are mainly distributed farther away from the surface (as highlighted in the black box).

In contrast, **Figure 9b_**(Fc30Pd10TAr) shows that Fe clusters are largely surrounded by cyclopentadienyl fragments, a behavior also observed in the Fc30TAr system. This comparison suggests that the presence of pyrazine or a higher nitrogen concentration enhances the reactivity of toluene toward Fe clusters, thereby altering the dominant carbon source in the vicinity of the catalyst.

Overall, these observations suggest that different carbon precursors contribute distinctly to the carbon supply near the catalyst. The stronger involvement of toluene-derived carbon species near the slab surface is further supported by the charge density profile of the mobile layer (**Figure 8c,** highlighted by dash box). Notably, a more pronounced charge depletion is observed in the Fc30Pd10TAr system (**Figure 9c**), particularly in the surface region highlighted by the box. This behavior suggests that the presence of pyridine in the network leads to relatively more neutral carbon species, which may contribute to defect formation during CNT growth, in agreement with experimental observations. Such charge depletion further suggests that it enhanced interactions between carbon species and the catalyst–substrate interface, influencing the growth mechanism and structural quality of the resulting CNTs.

Cluster analysis shows that the radius of gyration (**Rg**) of Fe clusters in the **Fc30Pz10TAr** system (**Figure 8d**) differs only slightly from those observed in **Fc30TAr** (**Figure 7d**) and **Fc30Pd10TAr** (**Figure 9d**). This suggests that nitrogen-containing precursors do not substantially alter the overall extent of Fe-cluster aggregation or compactness. Thus, the presence of nitrogen does not appear to significantly promote Fe atom aggregation or accelerate cluster formation under the investigated conditions. Instead, the observed differences suggest that nitrogen primarily modifies the interactions between precursor-derived carbon species and the Fe catalyst, potentially influencing the local chemical environment during the early stages of CNT formation. In particular, decomposition pathways are commonly associated with the formation of bamboo-like CNT structures, which are frequently observed in nitrogen-assisted CNT growth. These findings therefore support the idea that nitrogen incorporation may promote structural modifications in the catalyst that favor bamboo-shaped CNT formation. To further assess the consistency of this behavior, we performed statistical analysis based on nine independent MD simulations including above simulations as (SP0) for each system Fc30TAr, Fc30Pz10TAr and Fc30Pd10TAr respectively, in total 27 independent MD simulations.

Another important observation is related to the stability of the pyrazine molecule during the simulation. Unlike cyclopentadienyl and toluene precursors, which readily decompose into smaller fragments and carbon chains, the pyrazine molecule remains largely intact throughout the 4 ns simulation. This structural integration is highly compatible with the molecular dimensions of the precursor; in its relaxed ground state, the rigid, planar structure of pyrazine maintains a fixed distance of 2.7–2.8 Å between the two nitrogen atoms located at the 1,4-positions. This distance corresponds to opposite positions across the aromatic ring rather than a direct N–N bond, and such structural characteristics may facilitate nitrogen incorporation into the developing carbon framework during CNT growth.

Radial distribution function analysis (**Figure 8e**) further confirms the evolution of hydrogen species in the system. The absence of free proton species and the presence of a sharp peak corresponding to the characteristic H–H bond distance indicate the formation of molecular hydrogen ($H_2$). The comparison among different systems shows that hydrogen production is slightly higher in the Fc30TAr system, whereas the lowest hydrogen generation is observed in the Fc30Pd10TAr system. These differences suggest that the presence of nitrogen precursors or additional metal components modifies hydrocarbon decomposition pathways and hydrogen recombination dynamics. Overall, the results demonstrate that the introduction of pyrazine significantly alters Fe catalyst segregation behavior, carbon precursor decomposition, and the resulting catalyst particle sizes, which may ultimately influence CNT growth kinetics, diameter distribution, and the formation of nitrogen-related structural features such as bamboo-shaped CNTs. Moreover, the radial distribution function (RDF) analysis reveals significant interactions between nitrogen atoms and the carbon species originating from both toluene ($C_T$) and cyclopentadienyl ($C_p$) fragments. The RDF profile shows that nitrogen interacts with both carbon species; however, the intensity of the N–$C_p$ peak is noticeably higher than that of N–$C_T$. This suggests that nitrogen atoms exhibit a stronger affinity toward the cyclopentadienyl-derived carbon species compared to those originating from toluene. Additionally, as shown in **Figure 7e**, the RDF (Fc30Pz10TAr-red line) corresponding to Fe–Fe interactions display a relatively narrower Gaussian broadening, which can be attributed to the formation of more compact Fe clusters. This observation is consistent with the results obtained from the cluster analysis, where a lower radius of gyration also suggested enhanced compactness of the Fe clusters in the presence of pyrazine. Furthermore, the RDF analysis shows a prominent Fe–N interaction peak at approximately **2.15 Å**, indicating coordination between iron atoms and nitrogen species. Typically, the Fe–N bond length ranges from approximately **1.9 Å to 2.2 Å**, depending on the oxidation state, spin configuration, and coordination environment of the iron center. Shorter bond distances (around **1.9–2.0 Å**) are generally associated with higher oxidation states or strong metal–ligand back-bonding interactions, whereas longer distances (around **2.1–2.2 Å**) are commonly observed in high-spin iron species. The Fe–N distance observed in the present simulation therefore falls within the expected range, suggesting a realistic interaction between Fe clusters and nitrogen-containing species in the system.

In the Fc30Pd10TAr system (**Figure 9e**), the intensity of the N–CT peak is noticeably higher than that of the N–Cp peak, which is in contrast to the trend observed for pyrazine. This suggests that the molecular or atomic structure of the nitrogen precursor significantly influences the interaction with carbon species. Moreover, it contributes to charge redistribution, leading to enhanced charge depletion from the slab surface.

Pyridine has a relatively lower affinity toward Fe clusters, as evidenced by the radial distribution function (RDF), where the interaction distances fall within the range of approximately 2.8–4.5 Å. This weaker interaction further highlights how the nature of the nitrogen precursor governs catalyst–carbon interactions and ultimately impacts cluster evolution and CNT growth behavior.

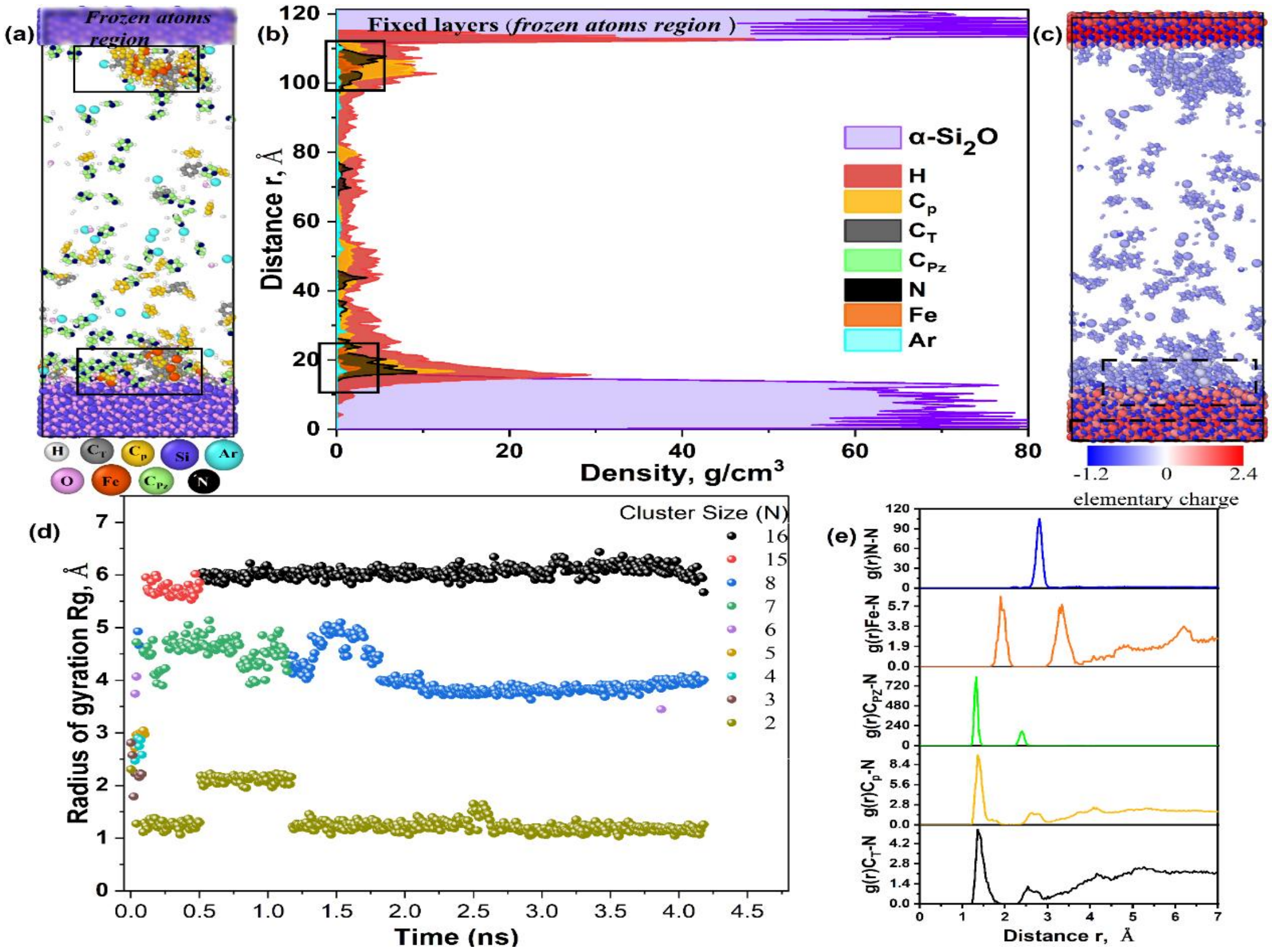

**Fig. 8:** Molecular dynamics post-processing results obtained from the first 5 ps of a 4 ns simulation at 760 °C (1033 K) for the Fc30Pz10TAr system: (a) atomic structures, (b) charge distribution profiles, (c) charge profiles, (d) cluster analysis results, and (e) radial distribution functions, *g(r)*, for N–C_T, N–$C_p$, N–$C_{Pz}$, Fe–N, and N–N pairs.

Overall, pyrazine exhibits slightly greater decomposition suggesting a modest difference in the availability and reactivity of precursor-derived species at the catalyst interface. These differences may influence carbon incorporation and the local chemical environment, potentially affecting the structural characteristics of the resulting CNTs. In comparison, pyridine shows relatively weaker interaction with the Fe clusters, which may lead to differences in the local distribution of reactive species. These trends are consistent with our experimental observations; however, the significance of these differences is further evaluated statistically in the following section.

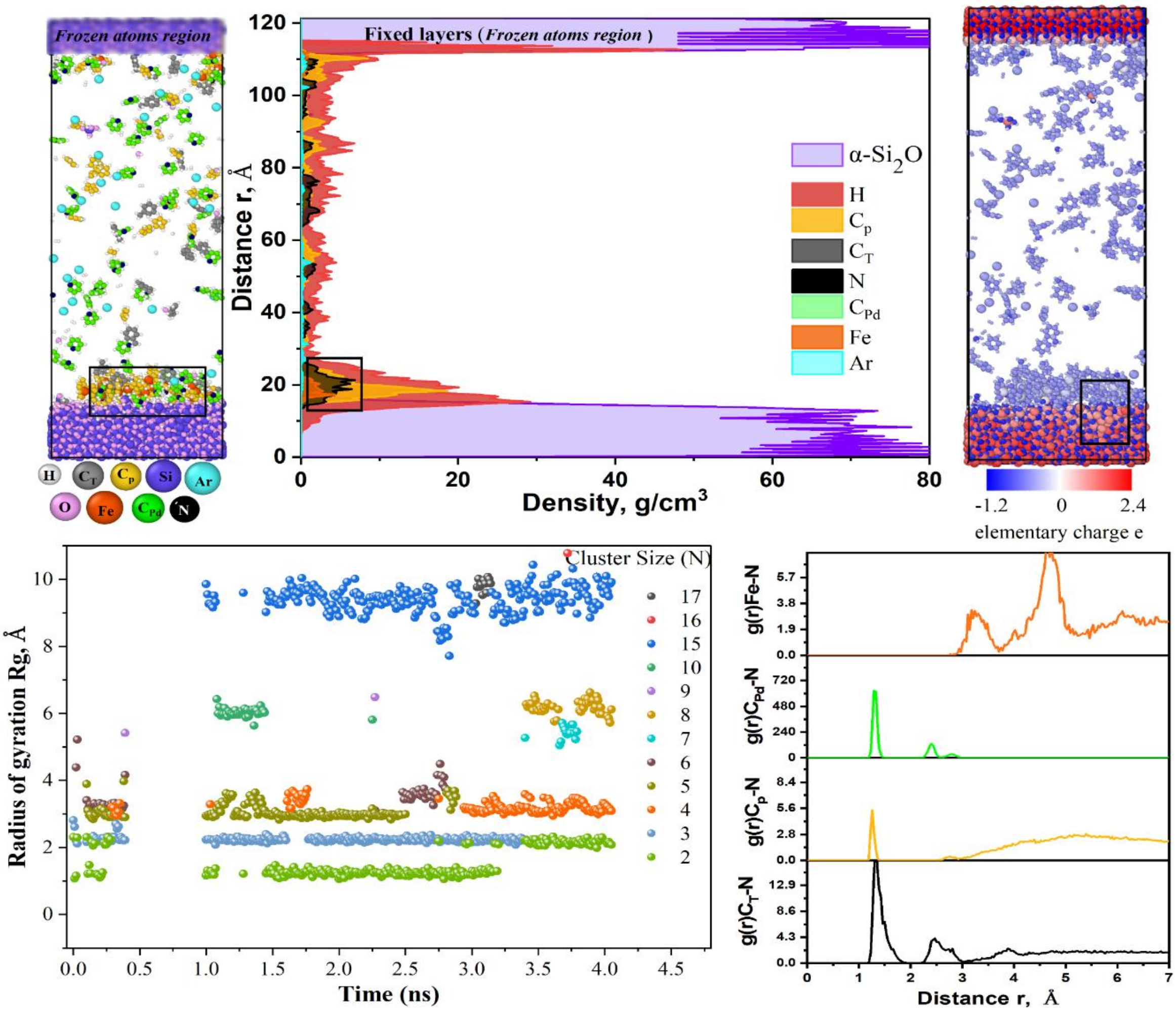

**Fig. 9:** Molecular dynamics post-processing results obtained from the first 5 ps of a 4 ns simulation at 760 °C (1033 K) for Fc30Pd10TAr. (a) Atomic structures, (b) charge distribution profiles, (c) charge profile, (d) cluster analysis results and (e) presents the radial distribution functions *g(r)* for N–$C_T$, N–$C_p$, N–$C_{Pd}$, Fe-N and N–N pairs.

### 3.2.4 Statistical analysis of precursor decomposition and catalyst interactions: Analysis of Fe-Cluster Evolution

Statistically significant differences were observed in the precursor decomposition and local catalyst interactions. The complete statistical analysis, including mean values, standard deviations, individual pairwise comparisons, and Holm-adjusted *p*-values, is provided in the Table1 (Raw data is provided in Supplementary Information) upto 4ns MD simulation. Pyrazine exhibited significantly greater decomposition than pyridine (Holm-adjusted $p = 3.97 \times 10^{-8}$), while the toluene IMF showed a significant difference between the Pz and Pd environments (Holm-adjusted $p = 0.0437$). Significant changes were also observed in the Fe–CT interaction for T–Pz and T–Pd comparisons (Holm-adjusted $p = 1.51 \times 10^{-12}$ and $1.84 \times 10^{-13}$, respectively), together with a significant difference observed in Fe–Cp distance for both Pz and Pd relative to T (Holm-adjusted $p$ =5.98 $10^{-7}$ and $1.09 \times 10^{-6}$ respectively). The Fe–$SiO_2$ equilibrium distance also increased significantly for both Pz and Pd relative to T (Holm-adjusted $p = 0.0025$). These statistically significant changes suggest that precursor chemistry primarily modifies molecular decomposition and the local Fe–molecule and Fe–$SiO_2$ interfacial interactions. Non-significant differences were observed for the T–Pd toluene IMF, Fe–N distance, direct Pz–Pd comparisons of Fe–CT, Fe–Cp and Fe–$SiO_2$, as well as Nfemax and Rg (Holm-adjusted $p > 0.05$). The detailed MD simulations-property analysis was first performed for a representative initial configuration, and the obtained descriptors were found to be consistent with the corresponding mean values across the nine independent MD trajectories for each system, supporting that the selected initial configurations reasonably represent the overall behaviour of the three systems comprising 27 independent simulations. Overall, the statistical analysis demonstrates that early-stage chemistry is primarily governed by precursor-dependent molecular decomposition and local Fe-cluster and Fe–$SiO_2$ interactions, with no statistically significant variation in the overall Fe-cluster aggregation behaviour across the investigated systems.

Table 1: Descriptive statistics and Holm-adjusted pairwise t-test parameters for T, Pz, and Pd systems across property metrics.

| Property / Metric | Group | Mean | SD | Comparison vs Group | Raw P-val | Holm-Adj P-val | Significant? |
|---|---|---|---|---|---|---|---|
| Toluen IMF | **T** | 37.67 | 6.60 | TvsPz<br>Tvs Pd | 0.0218<br>0.2222 | **0.0437**<br>0.2222 | **Yes**<br>No |
| | **Pz** | 30.00 | 6.20 | Pzvs Pd | 0.0024 | **0.0071** | **Yes** |
| | **Pd** | 42.00 | 7.83 | — | — | — | — |
| Fe-CT Equilibrium Distance, | **T** | 1.44 | 0.10 | T vs Pz<br>T vs Pd | $7.55 \times 10^{-13}$<br>$6.14\times 10^{-14}$ | $1.51\times 10^{-12}$<br>$1.84\times 10^{-13}$ | **Yes**<br>**Yes** |
| | **Pz** | 2.32 | 0.08 | Pz vs Pd | 0.1303 | 0.13033 | No |
| | **Pd** | 2.37 | 0.06 | — | — | — | — |
| Fe-Cp Equilibrium Distance (ED) | **T** | 3.72 | 0.51 | T vs Pz<br>T vs Pd | $2.99\times 10^{-7}$<br>$3.63\times 10^{-7}$ | $5.98\times 10^{-7}$<br>$1.09\times 10^{-6}$ | **Yes**<br>**Yes** |
| | **Pz** | 2.30 | 0.04 | Pz vs Pd | 0.1184 | 0.1184 | No |
| | **Pd** | 2.32 | 0.02 | — | — | — | — |
| Nitrogen precursor IMF | **Pz** | 74.22 | 1.72 | Pz vs Pd | $3.97\times 10^{-8}$ | $3.97\times 10^{-8}$ | **Yes** |
| | **Pd** | 88.11 | 3.92 | — | — | — | — |
| Cyclopentadienyl IMF | **T** | 16.22 | 3.49 | TvsPz<br>T vs Pd | 0.0285<br>0.0095 | 0.0570<br>0.0285 | No<br>**Yes** |
| | **Pz** | 20.89 | 4.65 | vs Pd | 0.6524 | 0.6524 | No |
| | **Pd** | 21.89 | 4.59 | — | — | — | — |
| Fe-SiO2 Equilibrium Distance (ED from density plot) | **T** | 3.39 | 0.66 | TvsPz<br>Tvs Pd | 0.0013<br>0.0003 | 0.0026<br>0.0009 | **Yes**<br>**Yes** |
| | **Pz** | 4.89 | 0.86 | Pzvs Pd | 0.8232 | 0.8232 | No |
| | **Pd** | 4.97 | 0.73 | — | — | — | — |
| Cluster analysis Nfemax | **T** | 16 | 3 | TvsPz<br>T vs Pd | 0.3714<br>0.4457 | 0.7427<br>0.4457 | No<br>No |
| | **Pz** | 14 | 5 | vs Pd | 0.8996 | 0.8996 | No |
| | **Pd** | 16 | 5 | — | — | — | — |
| Fe-N (from RDF) ED | **Pz** | 2.41 | 0.61 | Pzvs Pd | 0.5907 | 0.5907 | No |
| | **Pd** | 2.51 | 0.98 | — | — | — | — |
| Radius of gyration (Rg) | **T** | 6.80 | 1.89 | TvsPz<br>Tvs Pd | 0.4018<br>0.7879 | 0.8035<br>0.7879 | No<br>No |
| | **Pz** | 7.55 | 1.79 | vs Pd | 0.5639 | 0.5639 | No |
| | **Pd** | 7.04 | 1.85 | — | — | — | — |

The intact molecular fraction (IMF) is defined as $N_{IMF}$ (t)= ${}^{N}/_{N_0} \times 100$, where $N_0$ is the initial number of molecules and N is the number of intact molecules evaluated at 2.5 ns. $d_{Fe\text{-}x}$denotes the equilibrium Fe–X bond distance, where X= $C_T$, Cp; CT is Toluene's Carbon and $C_p$ represents cyclopentadienyl's carbon

## 4. Conclusion

This study demonstrates that nitrogen precursor chemistry strongly influences nitrogen incorporation and the structural characteristics of CNTs. Under comparable CVD conditions, both pyridine and pyrazine produce bamboo-like CNTs, but pyrazine results in more regular compartmentalization and higher nitrogen incorporation, whereas pyridine produces comparatively irregular and defect-rich structures. Raman and EDX analyses confirm these precursor-dependent differences.

Multiscale modeling shows that nitrogen addition does not substantially modify the reactor-scale flow and temperature fields. Instead, reactive molecular-dynamics simulations and statistical analysis reveal significant differences in precursor decomposition and local Fe–$C_T$, Fe–$C_p$, and Fe–$SiO_2$ interactions, with pyrazine exhibiting slightly greater decomposition than pyridine and enhancing toluene decomposition. In contrast, Fe-cluster size and compactness remain statistically insignificant, suggesting that the observed differences are not primarily associated with overall Fe-cluster aggregation.

Overall, the results suggest that precursor-dependent decomposition pathways and local catalyst–precursor and catalyst–substrate interactions play an important role in defining the early-stage chemical environment, which may contribute to the distinct CNT morphologies observed experimentally. This work highlights molecular precursor structure as an important parameter for controlling nitrogen incorporation and CNT structural characteristics.

**Author Contributions:**

**Narayan N. Som** Modelling, Investigation, Methodology, Visualizations, Data analysis, and Writing - original draft, review and editing; **Nowicki Lukasz** Data curation; Formal analysis; Investigation; Methodology; Visualization; Writing - original draft; and Writing - review & editing; **Gizewski Tomasz** Formal analysis; Investigation; Methodology; Software and Validation;; **Karolina Z. Milowska** Data analysis, Resources; Supervision, Validation, and Writing - review and editing**, Jakubowska Malgorzata** Supervision; Funding acquisition. **Lekawa-Raus Agnieszka** Conceptualization; Funding acquisition; Methodology; Project administration; Resources; Supervision; Validation and Writing - review & editing.

**Acknowledgements**

This work was supported by an Industry Cooperation funded by Infineon Technologies Austria AG during IPCEI Microelectronics. Ł.N., N.N.S., and A.L.-R. would also like to thank the Warsaw University of Technology Excellence Initiative (POSTDOC PW V edition PR-IDUB/374/201/210/2023 and Technologie Materialowe – 3 ADVANCED POB 1820/359/Z01/POB5/2021) for funding. N.N.S. and K.Z.M. gratefully acknowledge the Interdisciplinary Centre for Mathematical and Computational Modelling at the University of Warsaw, Poland (Grant No. G47-5) for providing computer facilities and technical support. N.N.S also gratefully acknowledges the Polish high-performance computing infrastructure PLGrid (HPC Center: ACK Cyfronet AGH) for providing computer facilities and support within the computational grant no. PLG/2025/019043. K.Z.M. acknowledges the technical and human support provided by the DIPC Supercomputing Center, Spain. K.Z.M would also like to thank the European Commission (Marie Skłodowska-Curie Cofund Programme; grant no. H2020-MSCA-COFUND-2020-101034228-WOLFRAM2) for supporting this research.

K.Z.M also would like to acknowledge the support from the Ramón y Cajal grant RYC2024-051436-I, funded by MICIU/AEI/10.13039/501100011033 and by ESF+.

**Data Availability Statement:** The datasets generated during and/or analyzed during the current study are available from the corresponding author on reasonable request

**Declaration of competing interest:** The authors declare that they have no known competing financial interests or personal relationships that could have appeared to influence the work reported in this paper.

## References

[1] W. Al-Hajri, Y. De Luna, N. Bensalah, Review on Recent Applications of Nitrogen-Doped Carbon Materials in CO 2 Capture and Energy Conversion and Storage, Energy Technol. 10 (2022). https://doi.org/10.1002/ente.202200498.

[2] W.H. Shin, H.M. Jeong, B.G. Kim, J.K. Kang, J.W. Choi, Nitrogen-Doped Multiwall Carbon Nanotubes for Lithium Storage with Extremely High Capacity, Nano Lett. 12 (2012) 2283–2288. https://doi.org/10.1021/nl3000908.

[3] W. Han, G. Liu, W. Seo, H. Lee, H. Chu, W. Yang, Nitrogen-doped chain-like carbon nanospheres with tunable interlayer distance for superior pseudocapacitance-dominated zinc- and potassium-ion storage, Carbon N. Y. 184 (2021) 534–543. https://doi.org/10.1016/j.carbon.2021.08.060.

[4] B.I. Orea-Calderón, C.G. Castillo, J.L. Varela-Caselis, J.C. Chacón-Torres, V.L. Medina-Llamas, L.A. Macclesh del Pino-Pérez, E. Muñoz-Sandoval, F. López-Urías, Nitrogen-doped carbon nanotubes with enhanced capacitance for electrochemical applications, Carbon Trends. 23 (2026) 100632. https://doi.org/10.1016/j.cartre.2026.100632.

[5] M. Christwardana, D. Dayanti, K. Khoirunnisa, G. Gunawan, D.S. Widodo, L. Suyati, A.Y. Maulana, J. Kim, Green hydrothermal engineering of nitrogen-doped carbon nanotubes using polyethylenimine for accelerated bioelectrocatalytic activity in yeast microbial fuel cells, J. Power Sources. 682 (2026) 240346. https://doi.org/10.1016/j.jpowsour.2026.240346.

[6] Y. Huang, M. Chen, A. Xie, Y. Wang, X. Xu, Recent Advances in Design and Fabrication of Nanocomposites for Electromagnetic Wave Shielding and Absorbing, Materials (Basel). 14 (2021) 4148. https://doi.org/10.3390/ma14154148.

[7] M. Shi, B. Zhang, X. Yan, J. Ma, X. He, N-doped carbon nanotubes as metal-free catalysts for PAA activation to degrade emerging pollutants: Exploration of reaction mechanisms and prediction of active sites, Environ. Res. 282 (2025) 121998. https://doi.org/10.1016/j.envres.2025.121998.

[8] F. Liu, S. Zheng, J. Wei, J. Liu, Y. Zhao, G. Zhang, Recent Progress in Nitrogen-Functionalized Carbon Materials for Supercapacitors: Synthesis, Properties, and Future Prospects, Chem. Rec. 26 (2026) e202500276. https://doi.org/10.1002/tcr.202500276.

[9] X. Chu, Y. Wang, J. Sun, Y. Wu, W. Jiang, B. Liu, C. Liu, G. Che, Y. Sun, X. Yang, Nitrogen-doped carbon nanotube encapsulated CoCu bimetallic alloy particles to promote efficient oxygen evolution reaction via electronic structure regulation, J. Colloid Interface Sci. 695 (2025) 137775. https://doi.org/10.1016/j.jcis.2025.137775.

[10] J. Briceño, D. Rosas-Medellín, I.L. Alonso-Lemus, B. Escobar, Nitrogen-doped carbon nanotubes with high-surface-area as electrocatalysts in alkaline media: Enhanced performance and life cycle assessment, Int. J. Hydrogen Energy. 141 (2025) 1153–1164. https://doi.org/10.1016/j.ijhydene.2025.02.048.

[11] A. Sharma, K. Dasgupta, S. Banerjee, A. Patwardhan, D. Srivastava, J.B. Joshi, In-situ nitrogen doping in carbon nanotubes using a fluidized bed reactor and hydrogen storage

behavior of the doped nanotubes, Int. J. Hydrogen Energy. 42 (2017) 10047–10056. https://doi.org/10.1016/j.ijhydene.2017.01.160.

[12] P. Korusenko, K. Kharisova, E. Knyazev, O. Levin, A. Vinogradov, E. Alekseeva, Surface Engineering of Multi-Walled Carbon Nanotubes via Ion-Beam Doping: Pyridinic and Pyrrolic Nitrogen Defect Formation, Appl. Sci. 13 (2023) 11057. https://doi.org/10.3390/app131911057.

[13] H. Zhou, Y. Yang, F. Qiao, R. Hong, H. Zhang, H. Chu, Effect of NH3 addition on the preparation of nitrogen-doped carbon nanomaterials by flame synthesis method, Carbon Lett. 34 (2024) 2343–2355. https://doi.org/10.1007/s42823-024-00760-y.

[14] Z. Miao, J. Meng, M. Liang, Z. Li, Y. Zhao, F. Wang, L. Xu, J. Mu, S. Zhuo, J. Zhou, In-situ CVD synthesis of Ni@N-CNTs/carbon paper electrode for electro-reduction of CO2, Carbon N. Y. 172 (2021) 324–333. https://doi.org/10.1016/j.carbon.2020.10.044.

[15] S. Boncel, S.W. Pattinson, V. Geiser, M.S.P. Shaffer, K.K.K. Koziol, En route to controlled catalytic CVD synthesis of densely packed and vertically aligned nitrogen-doped carbon nanotube arrays, Beilstein J. Nanotechnol. 5 (2014) 219–233. https://doi.org/10.3762/bjnano.5.24.

[16] T. Thurakitseree, A. Rattanachata, H. Nakajima, S. Phothiphiphit, S. Kuimalee, P. Suknet, Nitrogen-Doped Single-Walled Carbon Nanotubes by Floating-Catalyst CVD Process, Chem. Phys. 576 (2024) 112090. https://doi.org/10.1016/j.chemphys.2023.112090.

[17] E.N. Nxumalo, V.O. Nyamori, N.J. Coville, CVD synthesis of nitrogen doped carbon nanotubes using ferrocene/aniline mixtures, J. Organomet. Chem. 693 (2008) 2942–2948. https://doi.org/10.1016/j.jorganchem.2008.06.015.

[18] S. Boncel, S.W. Pattinson, V. Geiser, M.S.P. Shaffer, K.K.K. Koziol, En route to controlled catalytic CVD synthesis of densely packed and vertically aligned nitrogen-

doped carbon nanotube arrays, Beilstein J. Nanotechnol. 5 (2014) 219–233. https://doi.org/10.3762/bjnano.5.24.

[19] A. Hachimi, B. Merzougui, A. Hakeem, T. Laoui, G.M. Swain, Q. Chang, M. Shao, M.A. Atieh, Synthesis of Nitrogen-Doped Carbon Nanotubes Using Injection-Vertical Chemical Vapor Deposition: Effects of Synthesis Parameters on the Nitrogen Content, J. Nanomater. 2015 (2015) 453725. https://doi.org/10.1155/2015/453725.

[20] K. Chizari, A. Vena, L. Laurentius, U. Sundararaj, The effect of temperature on the morphology and chemical surface properties of nitrogen-doped carbon nanotubes, Carbon N. Y. 68 (2014) 369–379. https://doi.org/10.1016/j.carbon.2013.11.013.

[21] L.G. Bulusheva, A. V Okotrub, Y. V Fedoseeva, A.G. Kurenya, I.P. Asanov, O.Y. Vilkov, A.A. Koós, N. Grobert, Controlling pyridinic, pyrrolic, graphitic, and molecular nitrogen in multi-wall carbon nanotubes using precursors with different N/C ratios in aerosol assisted chemical vapor deposition, Phys. Chem. Chem. Phys. 17 (2015) 23741–23747. https://doi.org/10.1039/C5CP01981H.

[22] Y. Lu, Z. Zhu, D. Su, D. Wang, Z. Liu, R. Schlögl, Formation of bamboo-shape carbon nanotubes by controlled rapid decomposition of picric acid, Carbon N. Y. 42 (2004) 3199–3207. https://doi.org/10.1016/j.carbon.2004.08.003.

[23] S. van Dommele, A. Romero-Izquirdo, R. Brydson, K.P. de Jong, J.H. Bitter, Tuning nitrogen functionalities in catalytically grown nitrogen-containing carbon nanotubes, Carbon N. Y. 46 (2008) 138–148. https://doi.org/10.1016/j.carbon.2007.10.034.

[24] H. Chen, Y. Yang, Z. Hu, K. Huo, Y. Ma, Y. Chen, X. Wang, Y. Lu, Synergism of C 5 N Six-Membered Ring and Vapor−Liquid−Solid Growth of CN x Nanotubes with Pyridine Precursor, J. Phys. Chem. B. 110 (2006) 16422–16427. https://doi.org/10.1021/jp062216e.

[25] J.P. O'Byrne, Z. Li, S.L.T. Jones, P.G. Fleming, J.A. Larsson, M.A. Morris, J.D. Holmes,

Nitrogen-Doped Carbon Nanotubes: Growth, Mechanism and Structure, ChemPhysChem. 12 (2011) 2995–3001. https://doi.org/10.1002/cphc.201100454.

[26] Z. Su-Bing, S. Si-Xiu, Wei-Wei, X. Jun-Jian, L. Yan, The Preparation of Nitrogen-doped Carbon Nanotubes from Pyridine, Acta Physico-Chimica Sin. 20 (2004) 1320–1323. https://doi.org/10.3866/PKU.WHXB20041107.

[27] V. Eckert, E. Haubold, S. Oswald, S. Michel, C. Bellmann, P. Potapov, D. Wolf, S. Hampel, B. Büchner, M. Mertig, A. Leonhardt, Investigation of the surface properties of different highly aligned N-MWCNT carpets, Carbon N. Y. 141 (2019) 99–106. https://doi.org/10.1016/j.carbon.2018.09.035.

[28] S. Maldonado, S. Morin, K.J. Stevenson, Structure, composition, and chemical reactivity of carbon nanotubes by selective nitrogen doping, Carbon N. Y. 44 (2006) 1429–1437. https://doi.org/10.1016/j.carbon.2005.11.027.

[29] T. Sharifi, F. Nitze, H.R. Barzegar, C.-W. Tai, M. Mazurkiewicz, A. Malolepszy, L. Stobinski, T. Wågberg, Nitrogen doped multi walled carbon nanotubes produced by CVD-correlating XPS and Raman spectroscopy for the study of nitrogen inclusion, Carbon N. Y. 50 (2012) 3535–3541. https://doi.org/10.1016/j.carbon.2012.03.022.

[30] J. Liu, S. Webster, D.L. Carroll, Temperature and flow rate of NH3 effects on nitrogen content and doping environments of carbon nanotubes grown by injection CVD method, J. Phys. Chem. B. 109 (2005) 15769–15774. https://doi.org/10.1021/jp050123b.

[31] D. Hedman, B. McLean, C. Bichara, S. Maruyama, J.A. Larsson, F. Ding, Dynamics of growing carbon nanotube interfaces probed by machine learning-enabled molecular simulations, Nat. Commun. 15 (2024) 4076. https://doi.org/10.1038/s41467-024-47999-7.

[32] S. Sun, S. Maruyama, Y. Li, Chirality-Dependent Kinetics of Single-Walled Carbon Nanotubes from Machine-Learning Force Fields, J. Am. Chem. Soc. 147 (2025) 7103–

7112. https://doi.org/10.1021/jacs.4c18769.

[33] I. Kohata, R. Yoshikawa, K. Hisama, C. Bichara, K. Otsuka, S. Maruyama, Edge dynamics in iron-cluster catalyzed growth of single-walled carbon nanotubes revealed by molecular dynamics simulations based on a neural network potential, Carbon N. Y. 247 (2026) 120913. https://doi.org/10.1016/j.carbon.2025.120913.

[34] E.S. Penev, K. V Bets, J. Lei, B.I. Yakobson, Sulfur-Induced Shape Compliance Modulates Fe Catalyst Agglomeration in Carbon Nanotube Growth, Nano Lett. 26 (2026) 8385–8391. https://doi.org/10.1021/acs.nanolett.6c02124.

[35] Ł. Nowicki, N.N. Som, T. Gizewski, K.Z. Milowska, L. Al-Naimy, S. Lepak, J. Antonowicz, M. Szybowicz, A.B. Nowicka, M. Jakubowska, A. Lekawa-Raus, Role of early-stage chemistry and reactor environment in carbon nanotube formation by floating catalyst CVD, J. Mater. Chem. A. 14 (2026) 37382–37401. https://doi.org/10.1039/D6TA02276F.

[36] M. Li, J. Junnarkar, A.W.N. Sloan, E. Khabushev, A.V. Benavides-Figueroa, M.L. Gong, M. Garza, J.F. Khoury, D. Cavuto, S.M. Williams, others, Understanding the effect of transport phenomena in deep-injection floating catalyst chemical vapor deposition carbon nanotube synthesis, Carbon N. Y. 238 (2025) 120259. https://doi.org/10.1016/j.carbon.2025.120259.

[37] Y. Zhang, Z. He, K. Wu, Y. Niu, L. Zhao, S. Wang, H. Fu, J. Zhao, X. Li, M. Yan, Z. Yang, Scalable fabrication of carbon nanotube fibers via quad-channel FCCVD system, Carbon N. Y. (2026) 121571.

[38] A.P. Thompson, H.M. Aktulga, R. Berger, D.S. Bolintineanu, W.M. Brown, P.S. Crozier, P.J. In't Veld, others, LAMMPS-a flexible simulation tool for particle-based materials modeling at the atomic, meso, and continuum scales, Comput. Phys. Commun. 271 (2022) 108171.

[39] H.M. Aktulga, J.C. Fogarty, S.A. Pandit, A.Y. Grama, Parallel reactive molecular dynamics: Numerical methods and algorithmic techniques, Parallel Comput. 38 (2012) 245–259.

[40] C. Zou, A. Van Duin, Investigation of complex iron surface catalytic chemistry using the ReaxFF reactive force field method, Jom. 64 (2012) 1426–1437.

[41] Y. Wang, Y. Shi, Q. Sun, K. Lu, M. Kubo, J. Xu, Development of a transferable ReaxFF parameter set for carbon-and silicon-based solid systems, J. Phys. Chem. C. 124 (2020) 10007–10015.

[42] C.L. Chia, Classical and ReaxFF Molecular Dynamics Simulations of Fuel Additives at the Solid-Fluid Interface, The University of Manchester, 2019.

[43] S. Lepak-Kuc, A. Lekawa-Raus, M. Jakubowska, K. Koziol, Alternative Metallocenes in Floating Catalyst-CVD: Synthesis of Novel Carbon Nanostructures, Nanotechnol. Sci. Appl. (2025).

[44] E.N. Nxumalo, N.J. Coville, Nitrogen doped carbon nanotubes from organometallic compounds: a review, Materials (Basel). 3 (2010) 2141–2171.

[45] S.D. Mhlanga, E.N. Nxumalo, N.J. Coville, V. V Srinivasu, Nitrogen doping of CVD multiwalled carbon nanotubes: Observation of a large g-factor shift, Mater. Chem. Phys. 130 (2011) 1182–1186.

[46] S. Lepak-Kuc, S. Boncel, M. Szybowicz, A.B. Nowicka, I. Jozwik, K. Orlinski, T. Gizewski, K. Koziol, M. Jakubowska, A. Lekawa-Raus, The operational window of carbon nanotube electrical wires treated with strong acids and oxidants, Sci. Rep. 8 (2018) 14332. https://doi.org/10.1038/s41598-018-32663-0.

[47] Y. Fujimoto, Formation, structure, electronic, and transport properties of nitrogen defects in graphene and carbon nanotubes, Micromachines. 15 (2024) 1172.

[48] A. Lekawa-Raus, L. Kurzepa, G. Kozlowski, S.C. Hopkins, M. Wozniak, D. Lukawski,

B.A. Glowacki, K.K. Koziol, Influence of atmospheric water vapour on electrical performance of carbon nanotube fibres, Carbon N. Y. 87 (2015) 18–28. https://doi.org/10.1016/j.carbon.2015.02.018.

[49] F.-J. Zhang, B.-H. Zhou, X. Liu, Y. Song, X. Zuo, Molecular dynamics simulation of atomic hydrogen diffusion in strained amorphous silica, Chinese Phys. B. 29 (2020) 27101.

# Supporting Information

**Molecular Precursor Chemistry Shapes Early-Stage Reaction Pathways in Nitrogen-Doped Carbon Nanotube Synthesis**

*Narayan N. Som[1], Lukasz Nowicki[1], Tomasz Gizewski[2], Karolina Z. Milowska[3,4], Anna Wroblewska[5], Caghan Tuncer[6], Maciej Trzaskowski[1], Olga Iwasinska-Kowalska[7], Tomasz Plocinski[8], Mariusz Zdrojek[5],Krzysztof K. Koziol[6], Malgorzata Jakubowska[1,9], and Agnieszka Lekawa-Raus[1]**

[1] Centre for Advanced Materials and Technologies (CEZAMAT), Warsaw University of Technology, Warsaw, Poland;

[2] Lublin University of Technology, Faculty of Electrical Engineering and Computer Science, Lublin, Poland

[3]BCMaterials, Basque Center for Materials, Applications and Nanostructures, UPV/EHU Science Park, 48940 Leioa, Spain

[4]Ikerbasque, Basque Foundation for Science, Bilbao, Spain

[5] Faculty of Physics, Warsaw University of Technology, Warsaw, Poland

[6]Cranfield University, Cranfield, United Kingdom

[7]Faculty of Mechatronics, Warsaw University of Technology

8Faculty of Materials Science and Engineering

[9] Faculty of Mechanical and Industrial Engineering, Warsaw University of Technology, Warsaw, Poland

E-mail: agnieszka.raus@pw.edu.pl (A. Lekawa-Raus)

**Raman Analysis**

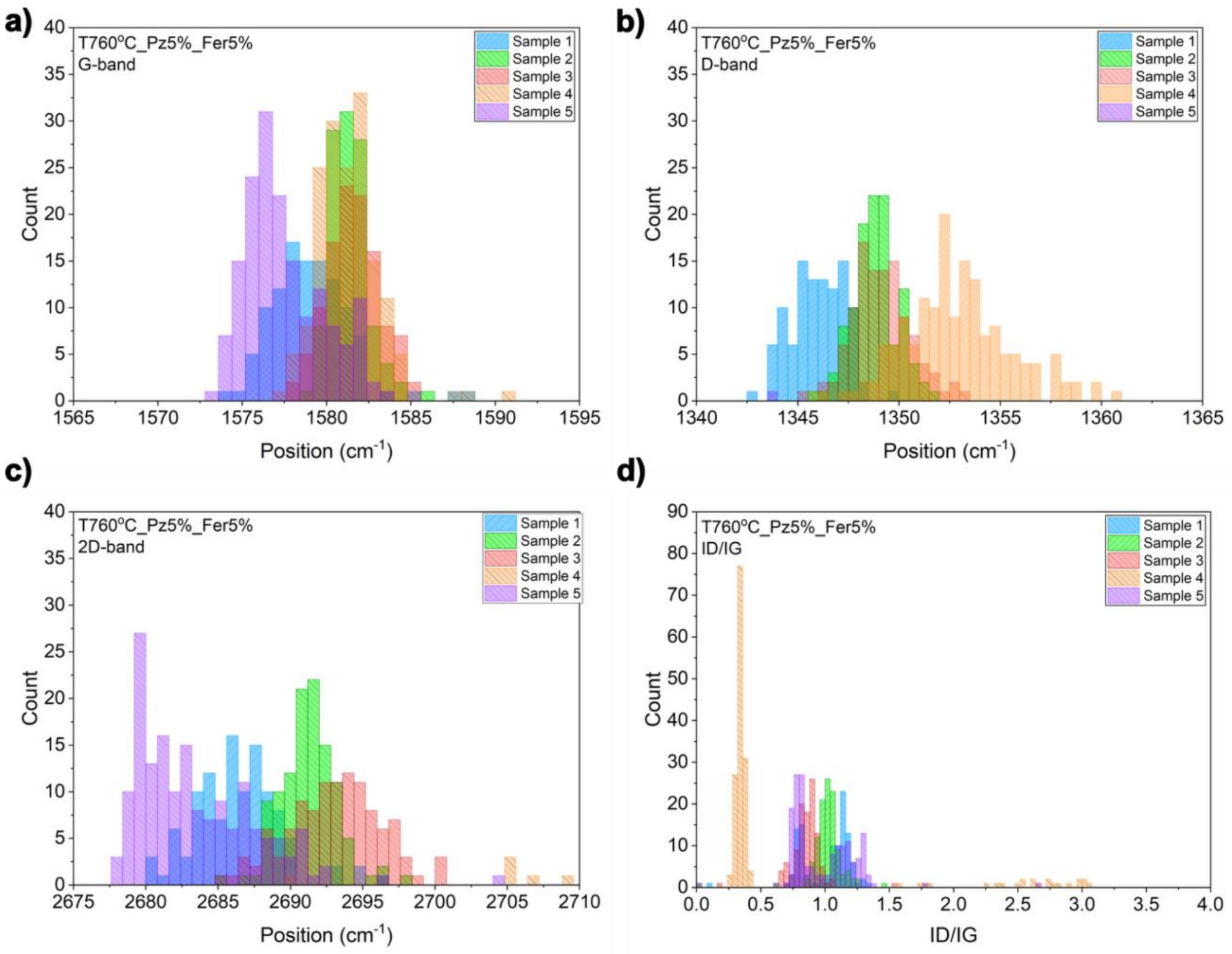

**Figure S1:** Raman shifts in the G, D, and 2D bands, together with changes in the $I_D/I_G$ ratio-for for 5% pyrazine precursor-based nitrogen-doped CNTs at 760 °C.

## S2 Numerical Analysis of Coupled Flow and Thermal Fields in a CVD Reactor

### S2.1 FEM modeling

In the present study, the CVD reactor is treated as a multiphysics system in which gas flow, heat transfer, and chemical reactions are strongly coupled. These phenomena are not considered independently, but are described within a unified modeling framework that reflects their mutual interactions. The gas flow field is not only responsible for transporting reactants and reaction products, but also plays an active role in shaping the local conditions under which deposition occurs. In particular, the velocity field determines precursor delivery to the reaction zone, residence time in the vicinity of the substrate, and the efficiency of product removal, all of which directly influence deposition uniformity [1].

The hydrodynamic description is inherently linked to the thermal model. In the gas phase, convection represents the dominant mechanism of energy transport, such that any change in flow structure is immediately reflected in the temperature distribution. Temperature, in turn, governs key aspects of the CVD process, including reaction rates, precursor decomposition pathways, and catalytic activity. Consequently, the flow and temperature fields form a coupled system, where neither field can be prescribed independently without compromising the physical consistency of the model.

Chemical reaction kinetics constitutes the third essential component of the methodology. Reaction rates are formulated as functions of local temperature and gas-phase composition, thereby coupling chemical transformations to both the thermal and hydrodynamic fields. Through reactant consumption and energy exchange associated with chemical reactions, the kinetic model influences local concentration fields and contributes to the overall energy balance. These effects feed back into the temperature field and, indirectly, into the flow behavior [1,2].

Based on these considerations, a coupled CFD–thermal–kinetic approach is adopted as the core methodological framework of this work. This integrated model enables a physically consistent description of transport and reaction phenomena within the CVD reactor and provides the basis for analyzing process stability, deposition uniformity, and sensitivity to operating parameters. Within the methodology presented here, Multiphysics modelling serves not only as a numerical tool, but as a structured means of interpreting and predicting reactor behaviour under varying process conditions.

A tubular CVD reactor (**Figure S2**), despite its apparent geometric simplicity, constitutes a system of significant physicochemical complexity. In the configuration considered in this study, the reactor consists of an outer shielding tube, inside which a process tube is placed coaxially. The process tube may be made of quartz ($SiO_2$), heat-resistant steel such as Kanthal (FeCrAl), or ceramic materials (e.g., $Al_2O_3$, SiC). The inner wall of the process tube constitutes the actual substrate surface (Γsub), on which the heterophase deposition reaction takes place.

The inter-wall space between the outer surface of the process tube and the inner surface of the shielding tube forms an annular reaction chamber (Ωann), through which the precursor gas mixture flows. In the case of a vertical configuration, the flow is driven by gravitational effects or inlet overpressure, whereas in the horizontal configuration forced flow dominates. The axial symmetry of the system allows the problem to be reduced to a two-dimensional domain formulated in cylindrical coordinates (r, z), where r denotes the radial coordinate measured from the reactor axis and z denotes the axial coordinate along the flow direction.

In the tubular CVD reactor, the process is governed by a combination of hydrodynamic, thermal, and mass-transport phenomena that evolve along the reactor axis. The gas flow develops into a fully established laminar velocity profile within the cylindrical channel,

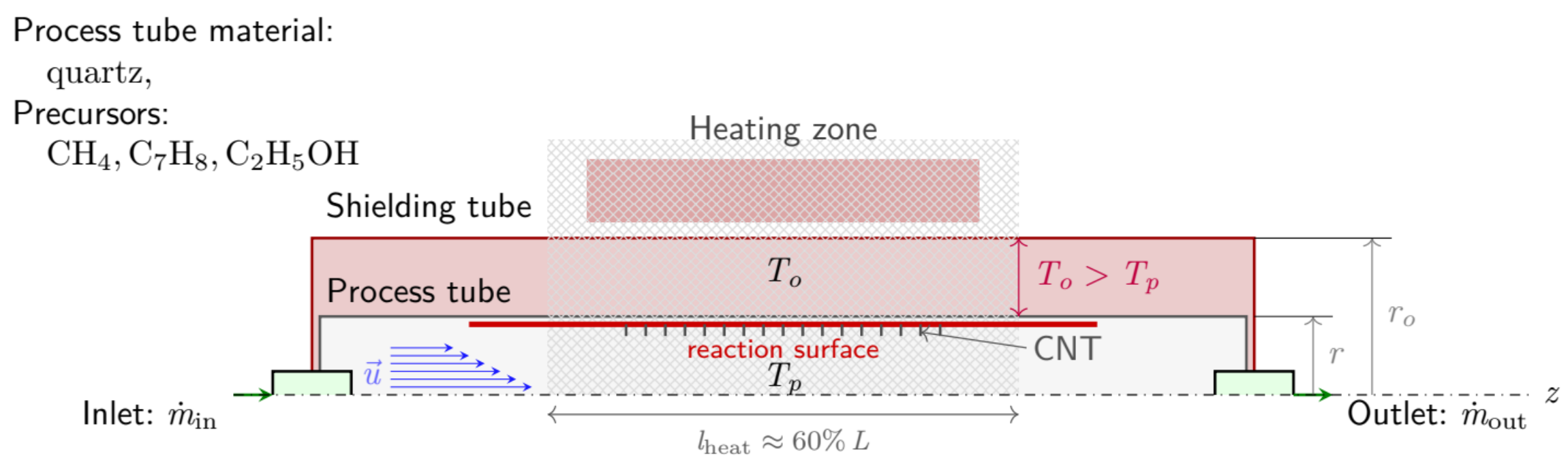

**Figure S2**: Schematic half-cross section of an axisymmetric tubular CVD reactor. The shielding tube surrounds the process tube (material: quartz, Kanthal, or ceramic). The heating zone (lheat ≈ 0.6L) covers the central part of the reactor; the gas flow ⃗u occurs inside a cylindrical domain of radius r. The shielding tube wall temperature To is higher than the process tube wall temperature $T_p$ ($T_o > T_p$). Carbon nanotubes (CNT) are deposited on the inner surface of the process tube (red line). The gas mixture enters

the reactor from the left (mass flow rate ṁ in), flows through the reaction zone, and exits on the right as exhaust products.

accompanied by the formation of a hydrodynamic boundary layer along the inner wall of the process tube, whose thickness δh(z) varies with the axial position. Simultaneously, a complex temperature field is established, characterized by radial thermal gradients in the gas phase ($\partial T/\partial r$), with maximum temperatures occurring near the indirectly heated process tube wall, as well as axial gradients ($\partial T/\partial z$) induced by the imposed heating profile. Under steady-state conditions, the process tube wall temperature $T_p(z)$ may remain lower than the shielding tube wall temperature $T_o(z)$ at the same axial location, owing to heat losses associated with gas-phase conduction and radiative transfer. Mass transport to the reaction surface proceeds through coupled convection and diffusion, delivering gas mixtures containing carbon precursors (methane $CH_4$, toluene $C_7H_8$, ethanol $C_2H_5OH$), a catalyst precursor (ferrocene $Fe(C_5H_5)_2$), sulfur-containing species (thiophene $C_4H_4S$), as well as carrier and modifying gases (argon Ar, hydrogen $H_2$, nitrogen $N_2$) to the curved, catalytically active surface where CNT growth occurs. The composition and relative proportions of the gas mixture constitute key technological parameters influencing nanotube morphology. Moreover, the curvature of the substrate surface alters local diffusive fluxes, modifying the effective thickness of the concentration boundary layer $\delta_c(r, z)$ compared to a planar geometry.

In the reactor configuration considered in this study, the CNT synthesis process takes place exclusively within the internal volume of the process tube, which serves as a sealed reaction chamber. The mixture of precursor and carrier gases is introduced directly into this space, bypassing the inter-tubular region between the process tube and the shielding tube. Such a flow arrangement, dictated by technological constraints and the requirement to ensure gas-tight operation, fundamentally alters the nature of the transport problem when compared to classical reactors operating with annular flow. The cylindrical geometry of the reaction channel imposes characteristic distributions of the physical fields and gives rise to coupled transport phenomena that are absent in systems based on planar geometries.

### S2.2 FEM Results

The internal volume of the process tube, characterized by a circular cross-section, represents a classical constant-area flow channel. Under steady laminar conditions, the axial velocity field assumes a parabolic profile, with a maximum at the reactor axis and a zero value at the inner wall of the process tube, which simultaneously serves as the reactive substrate surface. The no-slip condition at the gas–solid interface leads to the formation of a hydrodynamic boundary layer, whose thickness evolves along the reactor axis and directly affects the convective transport of reactants toward the active surface. Unlike annular-flow configurations, where velocity gradients are distributed between two confining walls, in the present geometry shear stresses and velocity gradients are concentrated exclusively near a single, reactive boundary.

A key feature of tubular reactor modeling is the presence of pronounced radial gradients of all physical fields. The temperature field exhibits a stable radial gradient directed from the heated process tube wall toward the cooler core of the gas stream, which may generate buoyancy-related effects and induce thermophoretic motion of aerosol particles or catalyst clusters. Simultaneously, species concentration fields are shaped by the competition between convective–diffusive transport from the bulk gas and consumption at the reactive wall, leading to precursor depletion in the near-wall region. These concentration gradients are inherently coupled to local reaction rates through their temperature dependence.

The parabolic velocity profile further governs axial transport by controlling residence time distributions within the reaction zone, thereby influencing precursor conversion and axial variations in process efficiency. In contrast to planar geometries, the cylindrical configuration introduces a geometric nonlinearity, as diffusive transport occurs across coaxial cylindrical surfaces whose area increases with radial distance. As a result, local diffusive fluxes depend not only on concentration gradients but also on the absolute radial position, rendering analytical treatment impractical and necessitating the use of advanced numerical methods for accurate process description.

In a tubular CVD reactor, where the reaction space takes the form of a cylindrical channel, heat and mass transport processes exhibit characteristic, strongly coupled behaviors that differ fundamentally from those observed in planar geometries. A central feature of such systems is the coexistence and competition of two transport mechanisms acting in orthogonal directions. Axial forced flow generates efficient convective transport along the reactor length, governing the residence time of the gas mixture in the heated zone and controlling the overall axial depletion of precursors. Shorter residence times reduce precursor consumption along the reactor but simultaneously limit the time available for surface reactions at a given axial position.

In contrast, participation of carbon precursors and doping gases, such as nitrogen, in heterogeneous reactions at the wall requires radial transport from the bulk gas toward the reactive surface. This radial transport is purely diffusive, driven by concentration gradients generated by surface consumption, and is characterized by a distinct time scale dependent on molecular diffusivity and the radial distance to the wall. The relative importance of axial convection and radial diffusion is conveniently described by the radial Péclet number, which under typical CVD conditions significantly exceeds unity. This indicates a transport regime dominated by axial convection, while the delivery of reactants to the wall is limited by comparatively slow radial diffusion, leading to the formation of a thin near-wall concentration boundary layer and diffusion-controlled reaction rates.

Heat transfer in high-temperature tubular reactors is likewise governed by multiple mechanisms. Radial heat conduction through the gas establishes temperature gradients between the heated process tube wall and the cooler flow core, while axial convection transports thermal energy along the reactor, contributing to axial temperature gradients, particularly near inlet and outlet regions. At temperatures typical of CNT synthesis, radiative heat transfer becomes a major contributor to the overall energy balance. Radiative exchange between the shielding tube and the process tube, strongly dependent on surface emissivity and cylindrical geometry, effectively transfers energy across the gas-filled gap. Neglecting radiative heat transfer under such conditions would result in a significant underestimation of the substrate temperature and, consequently, erroneous predictions of reaction kinetics.

A complete theoretical description of the processes occurring in a CVD reactor leads to a coupled system of partial differential equations, whose specific form depends on the adopted geometric representation. In the full three-dimensional (3D) formulation, without symmetry assumptions, the computational domain $\Omega_{3D}$ corresponds to the entire volume of the cylindrical reaction chamber, and the governing equations are expressed in the Cartesian coordinate system (x, y, z), with the z-axis aligned with the reactor axis.

$$\begin{cases} \nabla \cdot (\rho \vec{u}) = 0, & \text{(a) mass conservation,} \\ \rho (\vec{u} \cdot \nabla) \vec{u} = -\nabla p + \nabla \cdot \boldsymbol{\tau} + \rho \vec{g}, & \text{(b) momentum conservation (Navier-Stokes),} \\ \rho c_p (\vec{u} \cdot \nabla T) = \nabla \cdot (k \nabla T) + S_T, & \text{(c) energy conservation,} \\ \vec{u} \cdot \nabla (\rho Y_i) = \nabla \cdot (\rho D_i \nabla Y_i) + \dot{\omega}_i, & \text{(d) species transport.} \end{cases} \quad (S1)$$

The mathematical model consists of the conservation equations for mass, momentum, energy, and chemical species, which together define the coupled CFD–thermal–mass transport framework. These governing equations are summarized in Eqs. (S1)–(S4), forming the basis for the description of flow, heat transfer, and species transport in the gas phase.

In Eqs. (S1)–(S4), $\rho$ denotes the local gas mixture density, $\vec{u}$ the velocity vector, $p$ the static pressure, T the absolute temperature, and $Y_i$ the mass fraction of species i. The viscous stress tensor τ describes a Newtonian gas, $c_p$ is the specific heat capacity at constant pressure, $k$ the thermal conductivity, and $D_i$ the effective molecular diffusion coefficient. The terms $S_T$ and $\dot{\omega}_i$ represent volumetric heat sources (e.g. reaction heat) and net chemical source terms due to homogeneous reactions, respectively.

The governing system (S1)–(S4) is closed by the ideal gas equation of state, $p = \rho R_{mix} T$, where $R_{mix} = R/M_{mix}$ and $M_{mix}$denotes the mixture-averaged molar mass. Regardless of dimensionality, the system requires appropriate boundary conditions. At the inlet (z = 0), velocity profiles, temperature $T_{in}$, and species composition $Y_i$, it is prescribed. At the outlet (z = L), an open-boundary condition with zero normal gradients is typically applied for all variables except pressure. In the axisymmetric formulation, symmetry conditions are imposed along the reactor axis (r = 0), including zero radial gradients for scalar fields and vanishing radial velocity.

The most complex boundary condition is imposed at the reactive wall ($r = R_p$) in the 2D formulation or the corresponding surface in 3D). For the momentum equations, a no-slip condition is applied ($\vec{u}$= 0). For the energy equation, either a prescribed wall temperature $T_{wall}(\mathcal{Z})$ (Dirichlet condition) or a specified heat flux (Neumann condition) may be imposed, accounting for external heating and thermal losses. For the chemical species, a Neumann-type boundary condition is assumed, which relates the diffusive mass flux toward the wall to the rate of heterogeneous surface reaction $\dot{s}_i$:

$$-\rho D_i \left(\frac{\partial Y_i}{\partial n}\right)_{\text{wall}} = \dot{s}_i\left(Y_{i,\text{wall}}, T_{\text{wall}}\right).$$

The system is strongly coupled. The velocity field $\vec{u}$ obtained from the momentum equation (S1a) appears explicitly in the convective terms of the energy equation (S1c) and the species transport equations (S1d). The temperature field T, governed by the energy equation, influences the gas density ρ and viscosity μ in the momentum equations, as well as the reaction rate constants appearing in the source terms $\omega_i$ and $\dot{s}_i$. The species composition $Y_i$ determines the mixture-averaged molar mass and, consequently, the local gas density.

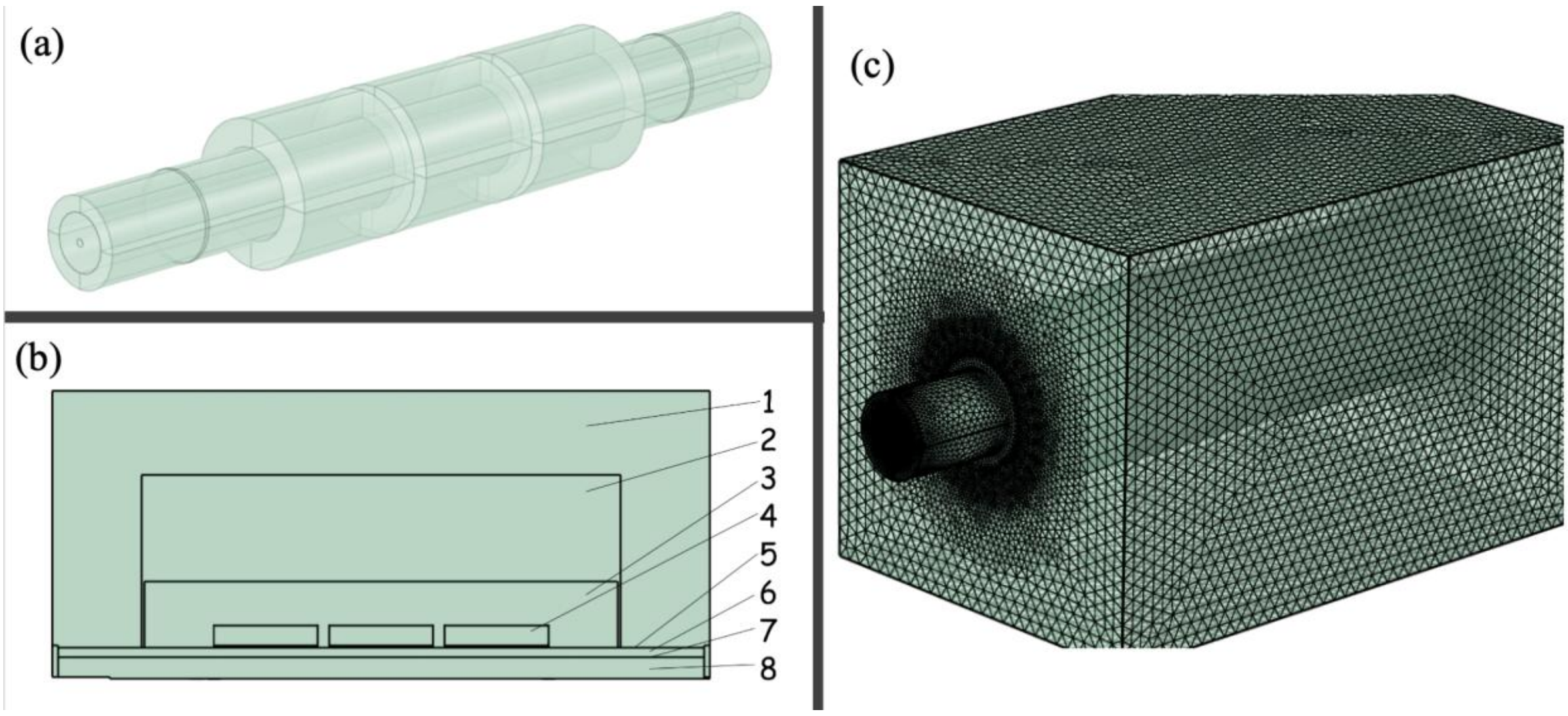

**Figure S3**: (a) Generalized 3D geometric model for CVD reactor simulations, excluding the surrounding ambient domain; (b) Two-dimensional axisymmetric model of the CVD furnace. Designation: (1) ambient air, (2) insulating ceramic, (3) insulating ceramic in the heating-element region, (4) heating elements, (5) shielding tube wall (kanthal), (6) interior of the shielding tube, (7) process tube wall, (8) interior of the process tube; and (c) Three-dimensional computational geometry of the tubular CVD furnace with generated numerical mesh, created using Gmsh and optimized for thermal and CVD analyses.

**Figure S3a** presents the geometry of the CVD furnace, highlighting the coaxial arrangement of the cylindrical components, including the process tube, the shielding tube, and the heating elements, which together form a symmetric configuration serving as the basis for subsequent numerical discretization. **Figure S3.** focuses on the arrangement of the tubes and heating elements alone, omitting the external ceramic insulation, thereby emphasizing the key cylindrical components responsible for thermal conditions and the gas-phase deposition process. This separation of perspectives facilitates a clearer interpretation of the reactor geometry and supports the preparation of the numerical model for further analysis.

The defined axisymmetric model enables the inclusion of heat conduction in solid materials, thermal interaction between individual zones, and heat exchange between the heating elements and the process region, providing a foundation for further numerical analyses.

**Figure S3c.** presents the three-dimensional computational geometry of the tubular CVD furnace together with the generated numerical mesh. The model consists of an outer insulating block and a coaxial arrangement of internal cylindrical components reflecting the actual furnace design, with the central process region surrounded by material layers of different thermal properties. The mesh was generated using the open-source software Gmsh and optimized for coupled thermal and CFD analyses.

An unstructured volumetric mesh based primarily on tetrahedral elements was applied, allowing accurate representation of the complex internal geometry. Local mesh refinement was introduced in regions of high physical relevance, such as near the process tube, shielding tube, and heating elements, while a coarser mesh was used in the ceramic insulation to reduce computational cost. The mesh exhibits satisfactory numerical quality, with acceptable element shape metrics and smooth size transitions, supporting stable and convergent simulations. In addition to the full 3D model, a two-dimensional axisymmetric formulation was also prepared as a computationally efficient alternative for preliminary and parametric studies.

Throughout the entire computational domain, comprising regions labeled from 1 to 8, the energy transport equation is solved in the form of the heat conduction equation. In the solid regions (1–7), such as insulating ceramics, heating elements, and tube walls, the absence of material motion is assumed, which leads to the classical Fourier heat conduction equation:

$$\rho c_p \frac{\partial T}{\partial t} = \nabla \cdot (k \nabla T) + Q, \qquad S2$$

where: $\rho$ denotes the material density, $c_p$ the specific heat capacity, T the temperature, $k$ the thermal conductivity, and Q a volumetric heat source (e.g. heat generation in the heating elements).

This equation applies both to the three-dimensional model formulated in a Cartesian coordinate system and to the two-dimensional axisymmetric model, with an appropriate interpretation of the divergence operator.

In region 8, corresponding to the interior of the process tube, in addition to the temperature field, the flow field of the process gas is also solved. Time-averaged Reynolds-averaged Navier–Stokes (RANS) equations are employed together with a $k$–ε turbulence model.

The momentum conservation equation takes the form:

$$\rho \frac{\partial \mathbf{u}}{\partial t} + \rho(\mathbf{u} \cdot \nabla)\mathbf{u} = \nabla \cdot [-p\mathbf{I} + \mathbf{K}] + \mathbf{F} + \rho\mathbf{g}, \qquad \text{S3}$$

where:

**u** is the velocity vector, p the pressure,

**I** the identity tensor,

**K** the viscous stress tensor,

**F** body forces other than gravity,

**g** the gravitational acceleration. The continuity equation is written as:

$$\frac{\partial \rho}{\partial t} + \nabla \cdot (\rho \mathbf{u}) = 0. \qquad \text{S4}$$

The viscous stress tensor K accounts for both molecular and turbulent viscosity:

$$\mathbf{K} = (\mu + \mu_T)(\nabla \mathbf{u} + (\nabla \mathbf{u})^T) - \frac{2}{3}(\mu + \mu_T)(\nabla \cdot \mathbf{u})\mathbf{I} - \frac{2}{3}\rho k \mathbf{I}, \qquad S5$$

where: μ is the dynamic viscosity, $\mu_T$ the turbulent viscosity, and k the turbulent kinetic energy.

The transport of turbulent kinetic energy is described by:

$$\rho \frac{\partial k}{\partial t} + \rho(\mathbf{u} \cdot \nabla)k = \nabla \cdot \left[\left(\mu + \frac{\mu_T}{\sigma_k}\right)\nabla k\right] + P_k - \rho\varepsilon, \qquad \text{S6}$$

whereas the transport equation for the turbulence dissipation rate is given by:

$$\rho \frac{\partial \varepsilon}{\partial t} + \rho(\mathbf{u} \cdot \nabla)\varepsilon = \nabla \cdot \left[\left(\mu + \frac{\mu_T}{\sigma_\varepsilon}\right)\nabla \varepsilon\right] + C_{\varepsilon 1}\frac{\varepsilon}{k}P_k - C_{\varepsilon 2}\rho\frac{\varepsilon^2}{k}. \qquad S7$$

The turbulent viscosity $\mu_T$ is computed as:

$$\mu_T = \rho C_\mu \frac{k^2}{\varepsilon}. \qquad \text{S8}$$

The turbulence production term $P_K$ is defined as:

$$P_k = \mu_T \left[\nabla \mathbf{u} : (\nabla \mathbf{u} + (\nabla \mathbf{u})^T) - \frac{2}{3}(\nabla \cdot \mathbf{u})^2\right] \qquad S9$$

In region 8, the energy equation assumes a convection-diffusion form:

$$\rho c_p \left(\frac{\partial T}{\partial t} + \mathbf{u} \cdot \nabla T\right) = \nabla \cdot [(k + k_T)\nabla T], \qquad S10$$

where $k_T$ denotes the turbulent thermal conductivity, which is related to the turbulent viscosity through the turbulent Prandtl number.

The objective of the numerical simulations was to qualitatively and quantitatively assess the influence of process gas composition, reactor tube material, and flow velocity on the velocity field, temperature field, and temperature gradients within the process region of a tubular CVD reactor. The analysis was performed in a transient (time-dependent) formulation over a total simulation time of 120 minutes, corresponding to typical durations of carbon deposition processes in CVD technologies.

The modeling was limited to coupled laminar flow and heat transfer phenomena. Chemical reaction kinetics and deposition processes were not included, as the purpose of the study was restricted to the hydrodynamic–thermal characterization of the reactor. The theoretical background and the formal description of the numerical model are presented in a separate part of the work.

The inlet boundary condition was defined by prescribing a mean flow velocity corresponding to a given volumetric flow rate, while a constant reference pressure condition was applied at the reactor outlet. Despite relatively low flow velocities and small tube diameters, turbulent flow was assumed in all cases due to gas dosing through a significantly smaller inlet capillary.

The temperature at the external boundaries of the computational domain was fixed at a constant ambient value:

$$T_{\text{ambient}} = 20°\text{C}.$$

The heating zone of the reactor was modeled as a volumetric heat source within the tube wall. The heating power was assumed to be time-dependent and followed an exponential function, with parameters determined experimentally based on the measured heating characteristics of the laboratory-scale reactor.

No temperature control algorithm was implemented in the model (i.e. no PID controller was applied). As a result, the heating process corresponds to an open-loop configuration, allowing the analysis of the natural thermal response of the system.

The simulations were carried out in a transient formulation over a total simulation time of

$$t_{\max} = 120 \text{ min.}$$

The results were evaluated at three characteristic time instants: t = 10 min, corresponding to the initial heating phase; t = 60 min, representing a quasi-steady regime; and t = 120 min, corresponding to the final stage of the process. For each of these time points, the gas velocity field (simulated for argon), the temperature field for all cases, and the temperature gradient within the process region were analyzed.

The numerical simulations included a gas scenario with nitrogen admixture. The objective of this part of the study was to assess the influence of the physical properties of the process gases on the velocity field, temperature distribution, and related thermal quantities, particularly under low-flow conditions. From a hydrodynamic perspective, the key physical properties of the process gases are density and molar mass, as they govern flow inertia and the level of kinetic energy transported by the gas stream.

**S3. Atomistic Decomposition Pathways of Ferrocene, Toluene, Pyrazine, and Pyridine**

We investigate the reaction mechanism occurring on the α-$SiO_2$ (101) slab to understand the formation of defective CNTs in the presence of nitrogen-containing precursors. The simulations are designed to reflect the experimental conditions that produced the best CNT growth, with 5% ferrocene at 760 °C. To systematically examine the effect of precursor concentration and composition, two model systems were considered containing 13 and 30 ferrocene molecules.

A fixed molecular basis of 13 ferrocene and 26 toluene molecules was used for all simulations. The reference system, Fc13TAr, represents a 1:2 ferrocene-to-toluene ratio. To investigate the effect of nitrogen-containing precursors, 13 pyrazine molecules were added for Fc13Pz5TAr, while 26 pyrazine molecules were used for Fc13Pz10TAr. Similarly, 26 pyridine molecules were added to construct Fc13Pd10TAr. These molecular numbers were selected to maintain a computationally tractable system while representing increasing nitrogen-precursor loading.

To investigate the influence of Fe precursor loading on catalyst evolution and CNT growth, analogous systems were constructed with 30 ferrocene molecules by replacing Fc13 with Fc30, while keeping the same relative percentage of Toluene precursor compositions. This modeling framework allows us to systematically evaluate how Fe concentration and nitrogen-containing precursors affect Fe cluster formation, surface reactions, and ultimately the development of nitrogen-related defects in carbon nanotubes.

### S3.1 Decomposition Behaviour of Ferrocene and Toluene with Nitrogen Precursors on the Quartz Surface

Our simulations reveal that ferrocene with toluene (Fc13T) at 760 °C, in the presence of argon gas, rapidly decomposes under pyrolytic conditions near the quartz surface, liberating Fe atoms that diffuse, cluster, and majorly anchor at oxygen sites on the substrate passivated with hydrogen. Toluene predominantly undergoes homolytic bond cleavage and hydrogen transfer reactions to generate small, highly mobile radicals such as methyl, benzyl, and short carbon fragments, which rapidly desorb or recombine within 1 ns without contributing to extended carbon framework formation. In the presence of hydrogen radicals, often existing as positively charged protons, additional hydrogen abstraction occurs from both the aromatic ring and the methyl side chain, further promoting radical generation and molecular fragmentation [3]. The decomposition pathway also produced methane $CH_4$, a $C_2H_2$ radical, and three methylidyne CH radicals, while three toluene derived radicals $C_6H_3CH_3$ were observed directly bonded to Fe atoms, indicating temporary metal-carbon interactions. Meanwhile, ferrocene underwent significant decomposition, with only five molecules remaining intact and the rest forming multiple linear carbon chains, including one C13 chain, one C10 chain, three C5 chains, and one C3 chain, highlighting its dominant role in supplying carbon for chain growth consistent with our previous work [4]. The lone hydrogen radical plays a significant role in toluene decomposition. Simultaneously, toluene suppresses secondary reactions between these radicals and ferrocene-derived fragments, limiting radical–radical coupling and chain growth and the largest chain observed was up to 10 carbon atoms. As a result, carbon species formed in the Fc13TAr system remain short-lived and spatially dispersed, with only minimal incorporation into cyclopentadienyl-derived chains, explaining the scarcity of long carbon chains observed. Moreover, the segregation of the Fe cluster limits to 3 to 4 atoms up to 2ns Molecular dynamics (MD) simulation (**See figure 6a and Figure S4**). The longest carbon chain observed consisted of 13 atoms, along with four separate 12 atom chains, among which one formed a 12 membered ring structure with an Fe atom positioned at the centre. In addition, carbon chains containing 11 and 9 atoms were identified. The number and length of these carbon chains increased progressively with simulation time, indicating continuous carbon aggregation. The formation of longer carbon chains may serve as precursors for graphitic layer development and could eventually contribute to the growth of multi walled carbon nanotube.

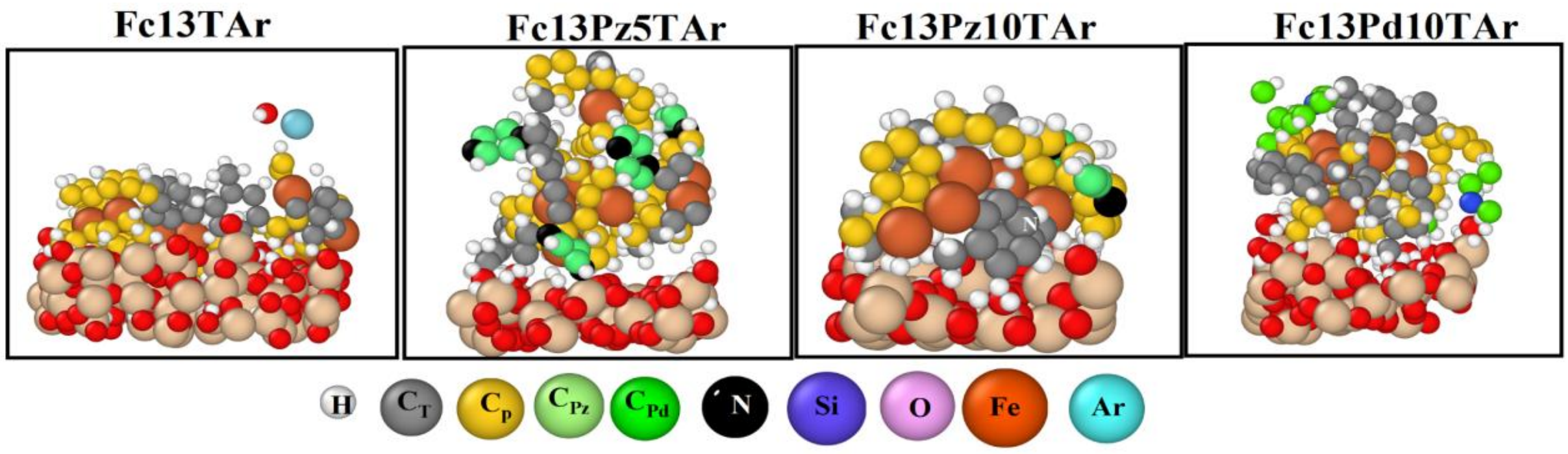

**Figure S4:** Zoomed-in view of Fe-cluster evolution during the molecular dynamics simulation, showing interactions between reactive radicals and the evolved Fe cluster at 2 ns. Atoms are color-coded as follows: hydrogen (H, white), toluene carbon ($C_T$, grey), cyclopentadienyl carbon ($C_{Cp}$, golden yellow), pyrazine carbon ($C_{Pz}$, pistachio green), pyridine carbon ($C_{Pd}$, electric green), nitrogen (N), silicon (Si, purple), oxygen (O, pink), iron (Fe, vibrant orange), and argon (Ar, cyan).

At low pyrazine concentration (Fc13Pz5TAr), nitrogen-containing fragments primarily adsorb in a planar configuration on the quartz surface or transiently coordinate with Fe atoms. Decomposition proceeds via ring opening and C–N bond cleavage, producing CN, $NH_3$, and small carbonaceous fragments is not observed till 2ns MD simulations. This behaviour aligns with earlier experimental and computational studies[5,6], suggesting that lower temperature and nitrogen concentrations favour substitutional N incorporation without severely disrupting carbon network formation.

Increasing the pyrazine concentration results in more frequent N–surface and N–Fe interactions, leading to enhanced fragmentation and a higher density of nitrogen-rich species near the surface. Compared to previous low-N systems, this higher concentration promotes the formation of Fe–N bonds, which are often associated with bamboo-like compartment formation in CNTs, as reported in experimental literature.

In contrast, Fc13 exhibits a comparatively different decomposition pathway. The single nitrogen atom in pyridine leads to less symmetric ring opening, with preferential cleavage adjacent to the N site. Relative to pyrazine, pyridine-derived fragments show reduced surface residence times and a greater tendency to desorb or diffuse into the gas phase, consistent with prior reports highlighting precursor-dependent nitrogen incorporation efficiencies.

The clustering behaviour of Fe atoms is found to be strongly correlated with both their segregation dynamics and the local chemical environment. With 13 ferrocene molecules and toluene in the presence of Argon (Fc13TAr), the initial nucleation of Fe clusters is observed at ~110 ps, where four dimers (two-atom clusters) are formed. As the simulation progresses to 140 ps, larger clusters consisting of 4, 3, and 2 atoms emerge. Notably, two of the clusters (sizes 2 and 4) remain stable and persist up to 2 ns of molecular dynamics (MD) simulation, indicating relatively slow coalescence and moderate surface mobility under nitrogen-free conditions.

In contrast, upon introducing low pyrazine concentration (Fc13Pz5TAr), a clear acceleration in cluster nucleation and growth is observed. Shortly after the formation of dimers, trimers appear more rapidly compared to the FcTAr system. By ~150 ps, clusters of 6 and 3 atoms are formed above the slab surface and remain stable until ~300 ps. Beyond this point, the 3-atom cluster coalesces with nearby Fe atoms to form a 4-atom cluster. This 4-atom cluster subsequently grows into a 5-atom cluster by ~1 ns, remaining relatively distant from the surface. Meanwhile, the 6-atom cluster segregates closer to the slab surface. Around 1.3 ns, this 6-atom cluster incorporates additional Fe atoms to form a 7-atom cluster, which remains stable in the same position up to 2 ns. These observations clearly indicate that the presence of nitrogen modifies Fe atom mobility and segregation, lowering surface segregation compared to the FcTAr system and stabilizing intermediate cluster sizes.

A further increase in nitrogen concentration (Fc13Pz10TAr), significantly enhances early-stage clustering kinetics. Clusters of 5, 4, and 2 atoms are detected as early as ~45 ps, followed by the formation of a 6-atom cluster at ~240 ps, while the smaller clusters persist. At ~800 ps, larger clusters of 9 and 4 atoms form, both positioned above the substrate surface. Interestingly, at ~1 ns, the 9-atom cluster fragments into 7- and 2-atom clusters, which remain stable until the end of the 2 ns simulation. The observed persistence of the 7-atom cluster in our simulations supports this concept. Furthermore, the $Fe_7$ cluster can serve as a stable core for the subsequent formation of $Fe_{13}$ clusters, which may promote a core–shell type growth pathway during nanoparticle evolution. When comparing Fc13Pz5TAr and Fc13Pz10TAr, we clearly observe that the latter exhibits more pronounced Fe cluster segregation. This trend is consistent with our experimental observations, where the growth of CNTs is found to be more favorable at 10% Pz than at 5% Pz.

For the Fc1310PzTAr system, up to ~600 ps the system predominantly contains four dimer clusters along with one trimer cluster. After ~650 ps, larger clusters consisting of 6 and 4 atoms begin to appear. By the end of the 2 ns simulation, mainly the 6-atom cluster together with several dimers remain segregated. These observations indicate that Fe cluster segregation is influenced not only by the concentration of nitrogen-containing precursors but also by the specific type of precursor used in the system.

At the same time, the formation of larger clusters ultimately depends on the availability of Fe atoms and the continuous supply of precursor species. In experimental growth conditions, the precursor is introduced gradually, which sustains cluster evolution over time. To better relate the simulations to these conditions, we therefore compare two different ferrocene loadings (13 and 30 molecules) in order to examine their influence on Fe cluster evolution and their potential role in carbon nanotube growth.

### S3.2 Statistical Analysis of Early-Stage Precursor Decomposition and Fe-Catalyst Interactions

To assess whether the observed differences in the atomistic descriptors are statistically significant, two-tailed Student's *t*-tests were performed using nine independent simulation trajectories for each system, followed by Holm correction for multiple comparisons. The mean values, standard deviations (SD), and Holm-adjusted *p*-values are summarized in Table 1 and raw data in Table S1 and Table S2. Within the Fc30TAr baseline environment (T), toluene exhibits a mean Intact Molecule Fraction (IMF) of 37.67% (SD = 6.60%), which decreases to 30% (SD = 6.21%) in the Fc30PzTAr environment (Pz). This decrease is statistically significant after Holm correction (Holm-adjusted $p = 0.0437$). In comparison, the Pz and Pd environments show a statistically significant difference in the remaining toluene fraction (Holm-adjusted $p = 0.0071$), with Pd retaining a higher IMF of 42 % (SD = 7.83%). The nitrogen precursors themselves show a pronounced difference in decomposition, with pyrazine retaining a mean IMF of 74.222% (SD = 1.716%) and pyridine (Pd) retaining 88.111% (SD = 3.919%) of their initial molecules. This difference is highly statistically significant (Holm-adjusted $p = 3.97 \times 10^{-8}$), indicating greater pyrazine decomposition under the investigated conditions.

The precursor-dependent decomposition is accompanied by significant changes in the local Fe–molecule interactions. The Fe–CT equilibrium distance increases from a mean value of 1.439 Å (SD = 0.099 Å) in T to 2.318 Å (SD = 0.084 Å) in Pz and 2.373 Å (SD = 0.063 Å) in Pd, with statistically significant differences relative to T (Holm-adjusted $p = 1.51 \times 10^{-12}$ and $1.84 \times 10^{-13}$, respectively), while the direct Pz–Pd comparison remains non-significant (Holm-adjusted $p = 0.1303$). Similarly, the Fe–Cp distance decreases from 3.721 Å (SD = 0.507 Å) in T to 2.300 Å (SD = 0.037 Å) and 2.323 Å (SD = 0.021 Å) in Pz and Pd, respectively, with significant differences relative to T (Holm-adjusted $p = 5.98 \times 10^{-7}$ for both), but no significant

Pz–Pd variation (Holm-adjusted p = 0.1185). The Fe–$SiO_2$ equilibrium distance also increases significantly from 3.394 Å (SD = 0.664 Å) in T to 4.887 Å (SD = 0.862 Å) and 4.972 Å (SD = 0.731 Å) for Pz and Pd, respectively (Holm-adjusted p = 0.0025 for T vs. Pz and p = 0.0010 for T vs. Pd), whereas the Pz–Pd comparison remains non-significant (Holm-adjusted p = 0.8232). In contrast, the Fe–N distance does not show a statistically significant difference between Pz and Pd (Holm-adjusted p = 0.5907).

The analysis shows that the largest Fe-cluster size (Nfemax)) does not exhibit a statistically significant difference among the toluene, pyrazine, and pyridine environments. The pristine toluene system (T) yielded a mean value of 16 (SD = 5), the pyrazine configuration (Pz) exhibited a mean value of 14 (SD = 3), and the pyridine configuration (Pd) showed a mean value of 16 (SD = 5). Pairwise evaluation via Student's t-test with a step-down Holm-Bonferroni correction confirmed no statistical variance across all comparisons, with Holm-adjusted p-values of 0.7427, 0.4457, and 0.8996 for T–Pz, T–Pd, and Pz–Pd, respectively. Similarly, the radius of gyration (($R_g$) remains statistically non-significant among all three networks. The mean values were established at 6.80 Å (SD = 1.89 Å) for T, 7.55 Å (SD = 1.79 Å) for Pz, and 7.04 Å (SD = 1.85 Å) for Pd, with no independent variance verified (Holm-adjusted p-values are 0.8035, 0.7879, and 0.8035 for T–Pz, T–Pd, and Pz–Pd, respectively. Overall, the statistical analysis indicates that precursor chemistry primarily affects molecular decomposition and the local Fe–molecule and (Fe–$SiO_2$) interactions, whereas the overall Fe-cluster aggregation and compactness remain statistically comparable under the investigated simulation condition.

**Table S1.** Key atomistic descriptors of precursor decomposition and Fe-catalyst interactions, including $N_{IMF}(4)$ for toluene and nitrogen precursors, equilibrium Fe–Cp and Fe–CT distances, and the Fe–$SiO_2$ equilibrium distance obtained from the highest peak of the density plot.

| Sample | Atomistic Mechanisms | | | | | | | | | | | | | |
|---|---|---|---|---|---|---|---|---|---|---|---|---|---|---|
| | Toluene remain intact ($N_{IMF}$ (4)) , % | | | Nitrogen precursor remain intact (NIMF (4)), % | | $d_{Fe–Cp}$ equilibrium distance, Å | | | $d_{Fe–CT}$ equilibrium distance from highest peak of RDF, Å | | | $Fe–SiO_2$ equilibrium distance from the highest peak of Density plot, Å | | |
| | T | Pz | Pd | Pz | Pd | T | Pz | Pd | T | Pz | Pd | T | Pz | Pd |
| SP0 | 35 | 23 | 38 | 75 | 92 | 3.72 | 2.33 | 2.34 | 1.43 | 2.32 | 2.3 | 4.10 | 4.50 | 4.40 |
| SP1 | 50 | 30 | 46 | 74 | 90 | 2.9 | 2.33 | 2.32 | 1.21 | 2.34 | 2.28 | 2.15 | 5.08 | 5.00 |
| SP2 | 38 | 38 | 42 | 76 | 94 | 3.05 | 2.3 | 2.29 | 1.52 | 2.33 | 2.46 | 2.95 | 3.45 | 6.50 |
| SP3 | 45 | 35 | 38 | 74 | 82 | 3.41 | 2.29 | 2.34 | 1.53 | 2.18 | 2.41 | 3.40 | 5.00 | 5.50 |
| SP4 | 34 | 31 | 61 | 73 | 90 | 4.1 | 2.31 | 2.33 | 1.44 | 2.35 | 2.42 | 3.30 | 4.50 | 4.50 |
| SP5 | 35 | 34 | 38 | 71 | 87 | 3.96 | 2.3 | 2.34 | 1.52 | 2.47 | 2.32 | 4.00 | 5.50 | 4.50 |
| SP6 | 31 | 26 | 42 | 73 | 83 | 4.19 | 2.35 | 2.29 | 1.41 | 2.34 | 2.42 | 3.20 | 5.95 | 4.50 |
| SP7 | 30 | 19 | 35 | 76 | 87 | 3.8 | 2.26 | 2.34 | 1.49 | 2.32 | 2.4 | 4.05 | 6.00 | 4.35 |
| SP8 | 41 | 34 | 38 | 76 | 88 | 4.36 | 2.23 | 2.32 | 1.4 | 2.21 | 2.35 | 4.00 | 4.00 | 5.50 |

**Table S2.** Comparison of key atomistic descriptors characterizing Fe-cluster evolution in different precursor environments: intact cyclopentadienyl fraction at 4 ns ($N_{IMF}$(4)), equilibrium Fe–N distance ($d_{Fe–N}$), largest Fe-cluster size, and radius of gyration (Rg).

| Sample | Atomistic Mechanisms | | | | | | | | | |
|---|---|---|---|---|---|---|---|---|---|---|
| | Cyclopentadienyl remaining intact (NIMF (4)), % | | | $d_{Fe–N}$, equilibrium distance, Å | | Largest Fe-cluster size, $NFe_{max}$, Atoms | | | Radius of gyration (Rg), Å | |
| | T | Pz | Pd | Pz | Pd | Pz | Pd | T | Pz | Pd | T |
| SP0 | 18 | 20 | 23 | 1.94 | 4.65 | 16 | 15 | 19 | 6.08 | 9.9 | 8.5 |
| SP1 | 14 | 23 | 12 | 1.94 | 1.99 | 25 | 12 | 16 | 10.27 | 4.86 | 6.77 |
| SP2 | 17 | 27 | 27 | 3.02 | 1.91 | 26 | 13 | 10 | 8.67 | 6.19 | 4.37 |
| SP3 | 20 | 22 | 20 | 1.94 | 1.89 | 13 | 22 | 11 | 8.23 | 10.17 | 4.96 |
| SP4 | 20 | 20 | 21 | 1.94 | 3.03 | 17 | 16 | 28 | 7.63 | 6.34 | 9.39 |
| SP5 | 10 | 15 | 20 | 3.42 | 1.92 | 16 | 16 | 15 | 9.46 | 6.85 | 6.68 |
| SP6 | 12 | 18 | 25 | 1.98 | 1.9 | 13 | 13 | 14 | 6.96 | 6.17 | 5.68 |
| SP7 | 17 | 15 | 22 | 3.04 | 3.34 | 10 | 11 | 13 | 5.63 | 7.43 | 5.47 |
| SP8 | 18 | 28 | 27 | 1.98 | 1.93 | 12 | 12 | 19 | 4.98 | 5.45 | 9.37 |

## Reference

S1. M. Li, J. Junnarkar, A.W.N. Sloan, E. Khabushev, A.V. Benavides-Figueroa, M.L. Gong, M. Garza, J.F. Khoury, D. Cavuto, S.M. Williams, others, Understanding the effect of transport phenomena in deep-injection floating catalyst chemical vapor deposition carbon nanotube synthesis, Carbon N. Y. 238 (2025) 120259. https://doi.org/10.1016/j.carbon.2025.120259.

S2. Zhang, Y., He, Z., Wu, K., Niu, Y., Zhao, L., Wang, S., Fu, H., Zhao, J., Li, X., Yan, M. and Yang, Z., 2026. Scalable fabrication of carbon nanotube fibers via quad-channel FCCVD system. *Carbon*, p.121571.

S3.Robaugh, D., and W. Tsang. "Mechanism and rate of hydrogen atom attack on toluene at high temperatures." *The Journal of Physical Chemistry* 90, no. 17 (1986): 4159-4163.

S4. Nowicki, Łukasz, Narayan N. Som, Tomasz Gizewski, Karolina Z. Milowska, Linda Al-Naimy, Sandra Lepak, Jerzy Antonowicz et al. "Role of early-stage chemistry and reactor environment in carbon nanotube formation by floating catalyst CVD." *Journal of Materials Chemistry A* (2026).

S5. Kiefer, J. H., R. D. Kern, H. Chen, J. Yao, and B. S. Jursic. "Dissociation and chain reaction in the pyrolysis of pyraine." In *Symposium (International) on Combustion*, vol. 26, no. 1, pp. 651-658. Elsevier, 1996.

S6. Zhu, Shuangfei, Shifa Cao, Zixuan Yang, Shufen Zheng, Shuhai Zhang, Ruijun Gou, and Guangrui Liu. "Quantum chemical modeling of the unimolecular decomposition reactions of energetic Azine compounds." *Computational and Theoretical Chemistry* 1248 (2025): 115212.